\documentclass{aa520}
\usepackage{epsfig,color,natbib}
\usepackage[dvips]{rotating}
\usepackage{xr2,afterpage}
\def\bibfiles{h3318bib}   

\def\aareferences{\longrefs=0  \bibliographystyle{aa}
            \bibliography{h3318aa,\bibfiles}}

\newcount\longrefs
\def\aap{\ifnum\longrefs=1 {Astron.\ Astrophys.}\else 
                           {A\hbox{\rm \&}A}\fi}
\def\aapr{\ifnum\longrefs=1 {Astron.\ Astrophys.\ Rev.}\else 
                            {A\hbox{\rm \&}AR}\fi}
\def\aaps{\ifnum\longrefs=1 {Astron.\ Astrophys.\ Suppl.}\else 
                            {A\hbox{\rm \&}AS}\fi}
\def\aj{\ifnum\longrefs=1 {Astron.\ J.}\else 
                          {AJ}\fi} 
\def\ao{\ifnum\longrefs=1 {Applied Optics}\else 
                           {Appl.\ Opt.}\fi} 
\def\aspcs{\ifnum\longrefs=1 {Astron.\ Soc.\ Pacific Conf. Series}\else 
                           {ASP Conf.\ Ser.}\fi} 
\def\apj{\ifnum\longrefs=1 {Astrophys.\ J.}\else 
                           {ApJ}\fi} 
\def\apjl{\ifnum\longrefs=1 {Astrophys.\ J. Lett.}\else 
                            {ApJ}\fi} 
\def\aplett{\ifnum\longrefs=1 {Astrophys.\ J. Lett.}\else 
                            {ApJ}\fi} 
\def\apjs{\ifnum\longrefs=1 {Astrophys.\ J. Suppl.}\else 
                            {ApJS}\fi}
\def\apss{\ifnum\longrefs=1 {Astrophys.\ and Space Science}\else 
                            {Ap\hbox{\rm \&}SS}\fi}
\def\araa{\ifnum\longrefs=1 {Ann.\ Rev.\ Astron.\ Astrophys.}\else 
                            {ARA\hbox{\rm \&}A}\fi}
\def\azh{\ifnum\longrefs=1 {Astronomicheskii Zhurnal}\else 
                            {Astron.\ Zhur.}\fi}
\def\baas{\ifnum\longrefs=1 {Bull.\ Am.\ Astron.\ Soc.}\else 
                            {BAAS}\fi}
\def\bain{\ifnum\longrefs=1 {Bull.\ Astronom.\ Institutes Netherlands}\else
                            {Bull.\ Astr.\ Inst.\ Neth.}\fi}
\def\gca{\ifnum\longrefs=1 {Geochim.\ Cosmochim.\ Acta}\else 
                           {Geochim.\ Cosmochim.\ Acta}\fi}
\def\grl{\ifnum\longrefs=1 {Geophys.\ Res.\ Lett.}\else 
                           {Geoph.\ Res.\ Lett.}\fi}
\def\iaucirc{\ifnum\longrefs=1 {IAU Circulars}\else 
                          {IAU Circ.}\fi}
\def\ip{\ifnum\longrefs=1 {in press}\else 
                          {in press}\fi}
\def\jgr{\ifnum\longrefs=1 {J.\ Geophys.\ Res.}\else 
                           {J.\ Geophys.\ Res.}\fi}  
\def\jrasc{\ifnum\longrefs=1 {J.\ Royal Astron.\ Soc.\ Canada}\else 
                           {JRAS Can.}\fi}  
\def\mnras{\ifnum\longrefs=1 {Mon.\ Not.\ Roy.\ Astron.\ Soc.}\else 
                             {MNRAS}\fi} 
\def\nat{\ifnum\longrefs=1 {Nature}\else 
                           {Nat}\fi}
\def\pasj{\ifnum\longrefs=1 {Pub.\ Astron.\ Soc.\ Japan}\else 
                            {PASJ}\fi} 
\def\pasp{\ifnum\longrefs=1 {Pub.\ Astron.\ Soc.\ Pacific}\else 
                            {PASP}\fi} 
\def\physscr{\ifnum\longrefs=1 {Physica Scripta}\else 
                            {Phys.\ Scrip.}\fi} 
\def\planss{\ifnum\longrefs=1 {Planetary \& Space Science}\else 
                            {Plan. \& Space Sci.}\fi} 
\def\procspie{\ifnum\longrefs=1 {Proc.\ SPIE}\else 
                            {Proc.\ SPIE}\fi} 
\def\qjras{\ifnum\longrefs=1 {Quarterly J.\ Royal Astron.\ Soc.}\else 
                            {QJRAS}\fi} 
\def\sa{\ifnum\longrefs=1 {Soviet Astron..}\else 
                               {Sov.\ Astron.}\fi}
\def\skytel{\ifnum\longrefs=1 {Sky \& Telescope}\else 
                            {Sky \& Tel.}\fi} 
\def\solphys{\ifnum\longrefs=1 {Solar Phys.}\else 
                               {Solar Phys.}\fi}
\def\ssr{\ifnum\longrefs=1 {Space Science Rev.}\else 
                               {Space\ Sci.\ Rev.}\fi}

\def\dutch{\def\refname{Referenties}\def\abstractname{Samenvatting}%
  \def\bibname{Bibliografie}\def\chaptername{Hoofdstuk}%
  \def\appendixname{Bijlage}\def\contentsname{Inhoudsopgave}%
  \def\listfigurename{Lijst van figuren}\def\listtablename{Lijst van tabellen}%
  \def\indexname{Index}\def\figurename{Figuur}\def\tablename{Tabel}%
  \def\partname{Deel}\def\enclname{Bijlage(n)}\def\ccname{Ter attentie van}%
  \def\headtoname{Aan}\def\headpagename{Pagina}%
  \def\today{\number\day\space\ifcase\month\or januari\or februari\or maart\or%
     april\or mei\or juni\or juli\or augustus\or september\or oktober\or%
     november\or december\fi \space\number\year}%
  \typeout{
              >>>>> use hlatex209 for Dutch hyphenation <<<<< 
         }}
\newcounter{onefig} \newcounter{fignumber}
\newcount\nocaptions \newcount\nofigures \newcount\figwidth
\newcount\viewgraphs
  \def\paper{}  \def\figlabel{} 
\long\def\nextfig#1{\setcounter{figure}{\value{fignumber}}
  \addtocounter{fignumber}{1}
  \ifnum \viewgraphs=1 \newpage \pagestyle{empty} \fi 
  \ifnum\value{onefig}=0 #1 \fi                 
  \ifnum\value{onefig}=\value{fignumber} #1 \fi}
\def\figwidths#1#2{\ifnum \nocaptions=1 #2mm \else #1mm \fi}  
\def\paper#1{}  
\long\def\plotfig#1#2{\ifnum \nofigures=1 \else #2 \fi}
\long\def\captiontext#1{\ifnum \nofigures=1 \raggedright \fi 
   \ifnum \nocaptions=1 \paper
     \ifnum \viewgraphs=0 
       \newline  \mbox{}\hrulefill\mbox{} \newline 
       \newline label:~\{\figlabel\} 
     \fi 
     \else \ifnum \nofigures=0 \fi 
   #1 \fi}

\newcount\panelwidth \newcount\panelheight 
\newcount\bxmin \newcount\bymin \newcount\bxmax \newcount\bymax
\newcount\tbxmin \newcount\tbymin
\newcount\tpanelwidth \newcount\tpanelheight \newcount\tpdif
\def\panelsize #1,#2;{\panelwidth=#1 \panelheight=#2}  
\def\setbb #1,#2;#3,#4;#5,#6;{
  \tbxmin=#1 \tbymin=#2    
  \bxmin=#3 \bymin=#4      
  \bxmax=#5 \bymax=#6}     
\def\barepanel #1{%
  \ifnum\panelheight=0 
    \tpdif=\bymax \advance\tpdif by -\bymin
    \multiply \tpdif by \panelwidth
    \tpanelheight=\tpdif
    \tpdif=\bxmax \advance\tpdif by -\bxmin
    \divide \tpanelheight by \tpdif
  \else \tpanelheight=\panelheight \fi
  \epsfig{file=#1,%
     bbllx=\bxmin bp,bblly=\bymin bp,bburx=\bxmax bp,bbury=\bymax bp,clip=,%
     width=\panelwidth mm,height=\tpanelheight mm}}
\def\labelypanel #1{
  \ifnum\panelheight=0 
    \tpdif=\bymax \advance\tpdif by -\bymin
    \multiply \tpdif by \panelwidth
    \tpanelheight=\tpdif
    \tpdif=\bxmax \advance\tpdif by -\bxmin
    \divide \tpanelheight by \tpdif
  \else \tpanelheight=\panelheight \fi
  \tpdif=\bxmax \advance\tpdif by -\tbxmin
  \tpanelwidth=\panelwidth \multiply \tpanelwidth by \tpdif
  \tpdif=\bxmax \advance\tpdif by -\bxmin
  \divide \tpanelwidth by \tpdif
  \epsfig{file=#1,%
    bbllx=\tbxmin bp,bblly=\bymin bp,bburx=\bxmax bp,bbury=\bymax bp,%
    clip=,width=\tpanelwidth mm,height=\tpanelheight mm}}
\def\labelxpanel #1{%
  \ifnum\panelheight=0 
    \tpdif=\bymax \advance\tpdif by -\bymin
    \multiply \tpdif by \panelwidth
    \tpanelheight=\tpdif
    \tpdif=\bxmax \advance\tpdif by -\bxmin
    \divide \tpanelheight by \tpdif
  \else \tpanelheight=\panelheight \fi
  \tpdif=\bymax \advance\tpdif by -\tbymin
  \multiply \tpanelheight by \tpdif
  \tpdif=\bymax \advance\tpdif by -\bymin
  \divide \tpanelheight by \tpdif
  \epsfig{file=#1,%
    bbllx=\bxmin bp,bblly=\tbymin bp,bburx=\bxmax bp,bbury=\bymax bp,%
    clip=,width=\panelwidth mm,height=\tpanelheight mm}}
\def\labelxypanel #1{%
  \ifnum\panelheight=0 
    \tpdif=\bymax \advance\tpdif by -\bymin
    \multiply \tpdif by \panelwidth
    \tpanelheight=\tpdif
    \tpdif=\bxmax \advance\tpdif by -\bxmin
    \divide \tpanelheight by \tpdif
  \else \tpanelheight=\panelheight \fi
  \tpdif=\bxmax \advance\tpdif by -\tbxmin
  \tpanelwidth=\panelwidth \multiply \tpanelwidth by \tpdif
  \tpdif=\bxmax \advance\tpdif by -\bxmin
  \divide \tpanelwidth by \tpdif 
  \tpdif=\bymax \advance\tpdif by -\tbymin 
  \multiply \tpanelheight by \tpdif
  \tpdif=\bymax \advance\tpdif by -\bymin
  \divide \tpanelheight by \tpdif
  \epsfig{file=#1,%
    bbllx=\tbxmin bp,bblly=\tbymin bp,bburx=\bxmax bp,bbury=\bymax bp,%
    clip=,width=\tpanelwidth mm,height=\tpanelheight mm}}

\def\CC{\par \vspace*{-2ex} \footnotesize \baselineskip=8pt \begin{verbatim}}

\long\def\startignore #1\stopignore{}   

\def\setlistparams{         
  \topsep=0.7ex                 
  \itemsep=0.7ex                
  \leftmargini=3ex}             
\setlistparams                  

\newcounter{alistindex}       

\newcounter{romenumnr}

\newlength{\minipagewidth}

\newsavebox{\boxcontent}
\newcommand{\ovalhead}[1]{
  \unitlength=1cm
  \sbox{\boxcontent}{\mbox{~~{#1}~~}}
  \begin{center}
    \ifdim\wd\boxcontent>6ex 
    \ifdim\wd\boxcontent<8cm 
    \begin{picture}(8,3) \thicklines     
      \put(4.0,0.8){\oval(8,1.6)} 
      \put(0.0,0.7){\parbox{8cm}{
         \begin{center} \usebox{\boxcontent} \end{center}}}
    \end{picture}
    \else \ifdim\wd\boxcontent<12cm 
    \begin{picture}(12,3) \thicklines     
        \put(6.0,0.8){\oval(12,1.6)} 
        \put(0.0,0.7){\parbox{12cm}{
           \begin{center} \usebox{\boxcontent} \end{center}}}
    \end{picture}
    \else
    \begin{picture}(16,3) \thicklines     
        \put(8.0,0.8){\oval(16,1.6)} 
        \put(0.0,0.7){\parbox{16cm}{
           \begin{center} \usebox{\boxcontent} \end{center}}}
    \end{picture}
    \fi \fi \fi
  \end{center}} 

\newcounter{headnr}            
\newcounter{subheadnr}[headnr]
\newcounter{subsubheadnr}[subheadnr]
\def\head #1\par{
  \stepcounter{headnr}                          
  \vspace{2ex} \noindent                        
  {\bf \theheadnr~~~~#1}\\[1ex] \noindent}      
\def\subhead #1\par{  
  \stepcounter{subheadnr}
  \vspace{1.3ex} \noindent
  {\bf \theheadnr.\arabic{subheadnr}~~~#1}\\[0.3ex] \noindent}
\def\subsubhead #1\par{
  \stepcounter{subsubheadnr}
  \vspace{1.0ex} \noindent
  {\bf \theheadnr.\arabic{subheadnr}.\arabic{subsubheadnr}~~~#1}\\ \noindent}

\font\dropfont= cmr12 scaled \magstep5
\def\dropcap#1#2{{\noindent
    \setbox0\hbox{\dropfont #1}\setbox1\hbox{#2}\setbox2\hbox{(}%
    \count0=\ht0\advance\count0 by\dp0\count1\baselineskip
    \advance\count0 by-\ht1\advance\count0by\ht2
    \dimen1=.5ex\advance\count0by\dimen1\divide\count0 by\count1
    \advance\count0 by1\dimen0\wd0
    \advance\dimen0 by.25em\dimen1=\ht0\advance\dimen1 by-\ht1
    \global\hangindent\dimen0\global\hangafter-\count0
    \hskip-\dimen0\setbox0\hbox to\dimen0{\raise-\dimen1\box0\hss}%
    \dp0=0in\ht0=0in\box0}#2}

\def\level #1 #2#3#4{$#1 \: ^{#2} \mbox{#3} ^{#4}$}   

\def\kms{\hbox{km$\;$s$^{-1}$}}

\def\mathstacksym#1#2#3#4#5{\def#1{\mathrel{\hbox to 0pt{\lower 
    #5\hbox{#3}\hss} \raise #4\hbox{#2}}}}

\mathstacksym\lta{$<$}{$\sim$}{1.5pt}{3.5pt} 
\mathstacksym\gta{$>$}{$\sim$}{1.5pt}{3.5pt} 
\mathstacksym\lrarrow{$\leftarrow$}{$\rightarrow$}{2pt}{1pt} 
\mathstacksym\lessgreat{$>$}{$<$}{3pt}{3pt} 

\bibpunct{(}{)}{;}{a}{}{,}
\newcommand{\ang}{$\rm \AA$}

\newcommand{\Msun}{M$_{\odot}$}
\newcommand{\Rsun}{R$_{\odot}$}
\newcommand{\Lsun}{L$_{\odot}$}
\newcommand{\be}{\begin{equation}}
\newcommand{\ee}{\end{equation}}
\newcommand{\bee}{\begin{eqnarray}}
\newcommand{\ad}{$\theta_d$}
\newcommand{\vt}{$\xi_t$}
\newcommand{\cc}{$\mathrm{^{12}C/^{13}C}$}

\newcommand{\ene}{\end{eqnarray}}
\newcommand{\teff}{T$_{\mathrm{eff}}$}
\newcommand{\mic}{$\mu$m}

\begin{document}

\title{ISO-SWS calibration and the accurate modelling of cool-star
atmospheres \thanks{Based on observations with ISO, an ESA project with
instruments funded by ESA Member States (especially the PI countries France,
Germany, the Netherlands and the United Kingdom) and with the participation of
ISAS and NASA.}}
\subtitle{III.\ A0 -- G2 stars: APPENDIX}

\author{L.~Decin\inst{1}\thanks{\emph{Postdoctoral Fellow of the Fund for
Scientific Research, Flanders}}  \and
B. Vandenbussche\inst{1} \and
C.~Waelkens\inst{1}
K.~Eriksson\inst{2} \and
B.~Gustafsson\inst{2} \and
B.~Plez\inst{3} \and
A.J.~Sauval\inst{4}}

\offprints{L.\ Decin, e-mail: Leen.Decin@ster.kuleuven.ac.be}

\institute{Instituut voor Sterrenkunde, KULeuven, Celestijnenlaan 200B, B-3001
    Leuven, Belgium
\and
    Institute for Astronomy and Space Physics, Box 515, S-75120 Uppsala, Sweden
\and
    GRAAL - CC72, Universit\'{e} de Montpellier II, F-34095 Montpellier Cedex 5,
France
\and
    Observatoire Royal de Belgique, Avenue Circulaire 3, B-1180 Bruxelles,
    Belgium
}

\date{Received data; accepted date}

\abstract{Vega, Sirius, $\beta$ Leo, $\alpha$ Car and $\alpha$ Cen A
belong to a  sample of twenty stellar sources used for the
calibration of the detectors of the Short-Wavelength Spectrometer on
board the Infrared Space Observatory (ISO-SWS). While general
problems with the calibration and with the theoretical modelling of
these stars are reported in \citet{Decin2000b}, each of these stars
is discussed individually in this paper. As demonstrated in
\citet{Decin2000b}, it is not possible to deduce the effective
temperature, the gravity and the chemical composition from the ISO-SWS
spectra of these stars. But since ISO-SWS is absolutely calibrated,
the angular diameter (\ad) of these stellar sources can be deduced from
their ISO-SWS spectra, which consequently yields the stellar radius (R),
the gravity-inferred mass (M$_g$) and the luminosity (L) for these
stars. For Vega, we obtained \ad$ = 3.35 \pm 0.20$\,mas, R$ = 2.79 \pm
0.17$\,\Rsun, M$_g = 2.54 \pm 1.21$\,\Msun\ and L$ = 61 \pm 9$\,\Lsun;
for Sirius \ad$ = 6.17 \pm 0.38$\,mas, R$ = 1.75 \pm
0.11$\,\Rsun, M$_g = 2.22 \pm 1.06$\,\Msun\ and L$ = 29 \pm 6$\,\Lsun;
for $\beta$ Leo \ad$ = 1.47 \pm 0.09$\,mas, R$ = 1.75 \pm
0.11$\,\Rsun, M$_g = 1.78 \pm 0.46$\,\Msun\ and L$ = 15 \pm 2$\,\Lsun;
for $\alpha$ Car \ad$ = 7.22 \pm 0.42$\,mas, R$ = 74.39 \pm
5.76$\,\Rsun, M$_g = 12.80^{+24.95}_{-6.35}$\,\Msun\ and L$ = 14573 \pm
2268$\,\Lsun\ and for $\alpha$ Cen A \ad$ = 8.80 \pm 0.51$\,mas, R$ = 1.27 \pm
0.08$\,\Rsun, M$_g = 1.35 \pm 0.22$\,\Msun\ and L$ = 1.7 \pm 0.2$\,\Lsun.
These deduced parameters are confronted with other published values and the
goodness-of-fit between observed ISO-SWS data and the
corresponding synthetic spectrum is discussed.
}

\maketitle
\markboth{L.\ Decin et al.: ISO-SWS and modelling of cool stars}{L.\ Decin et
al.: ISO-SWS and modelling of cool stars}

\appendix
\section{Comparison between different ISO-SWS and synthetic
spectra {\rm{(coloured plots)}}}

In this section, Fig.\ \ref{acenvers} -- Fig.\ \ref{acar}, Fig.\
\ref{bleovers} -- Fig.\ \ref{vega} of the accompanying article are
plotted in colour in order to better distinguish the different
observational or synthetic spectra.

\begin{figure}
\begin{center}
\resizebox{0.48\textwidth}{!}{\rotatebox{90}{\includegraphics{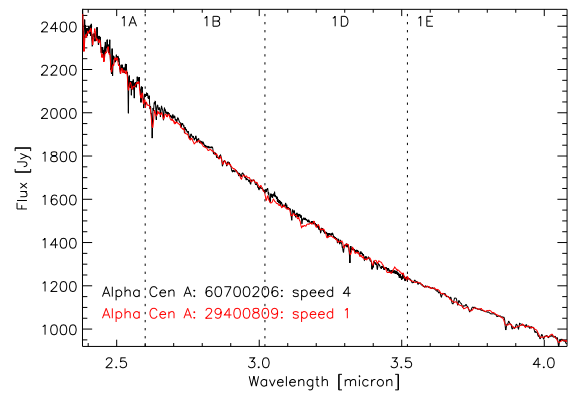}}}
\vspace*{-3.ex}
\caption{\label{acenverscol}Comparison between the AOT01 speed-4
observation of \boldmath {\bf{$\alpha$ Cen A}} \unboldmath (revolution
607) and the speed-1
observation (revolution 294). The data of the speed-1 observation
have been multiplied by a factor 1.16.}
\end{center}
\end{figure}

\begin{figure}
\begin{center}
\resizebox{0.5\textwidth}{!}{\rotatebox{90}{\includegraphics{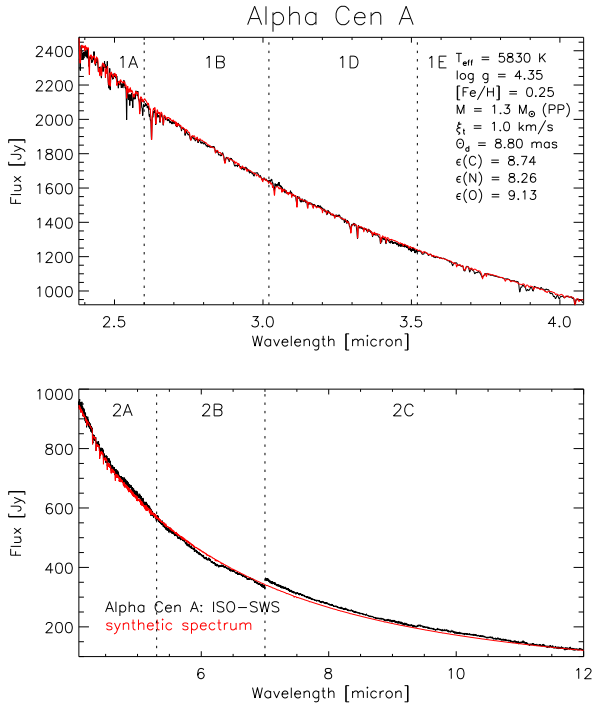}}}
\vspace*{-7ex}
\caption{\label{acencol} Comparison between band 1 and band 2 of the
ISO-SWS data of \boldmath {\bf{$\alpha$ Cen A}} \unboldmath (black)
and the synthetic spectrum
(red) with stellar parameters \teff\ = 5830\,K, $\log$ g = 4.35, M
= 1.3\,\Msun, [Fe/H] = 0.25, \vt\ = 1.0\,\kms, $\varepsilon$(C)
= 8.74, $\varepsilon$(N) = 8.26, $\varepsilon$(O) = 9.13 and \ad\
= 8.80\,mas.}
\end{center}
\end{figure}

\begin{figure}
\begin{center}
\resizebox{0.5\textwidth}{!}{\rotatebox{90}{\includegraphics{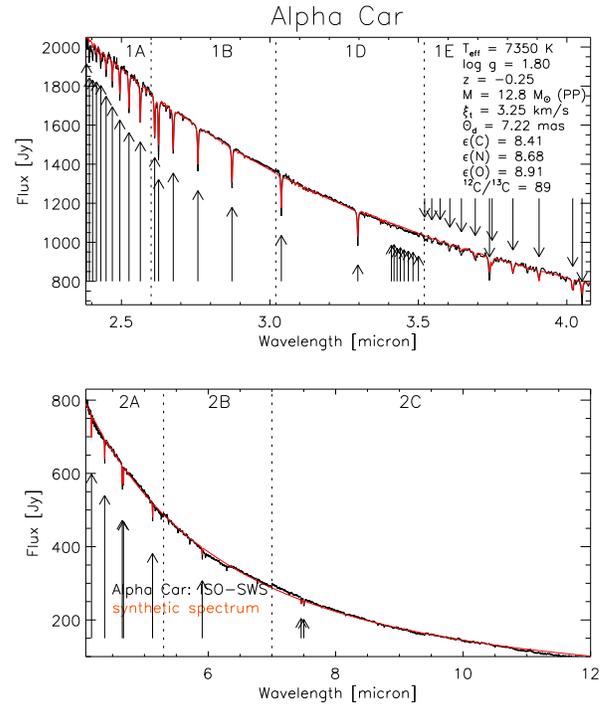}}}
\vspace*{-7ex}
\caption{\label{acarcol} Comparison between band 1 and band 2 of the
ISO-SWS data of \boldmath {\bf{$\alpha$ Car}} \unboldmath (black) and
the synthetic spectrum
(red) with stellar parameters \teff\ = 7350\,K, $\log$ g = 1.80, M
= 12.8\,\Msun, [Fe/H] = $-0.25$, \vt\ = 3.25\,\kms,
$\varepsilon$(C) = 8.41, $\varepsilon$(N) = 8.68, $\varepsilon$(O)
= 8.91 and \ad\ = 7.22\,mas. Hydrogen lines are indicated by
arrows.}
\end{center}
\end{figure}

\begin{figure}[h!]
\begin{center}
\resizebox{0.5\textwidth}{!}{\rotatebox{90}{\includegraphics{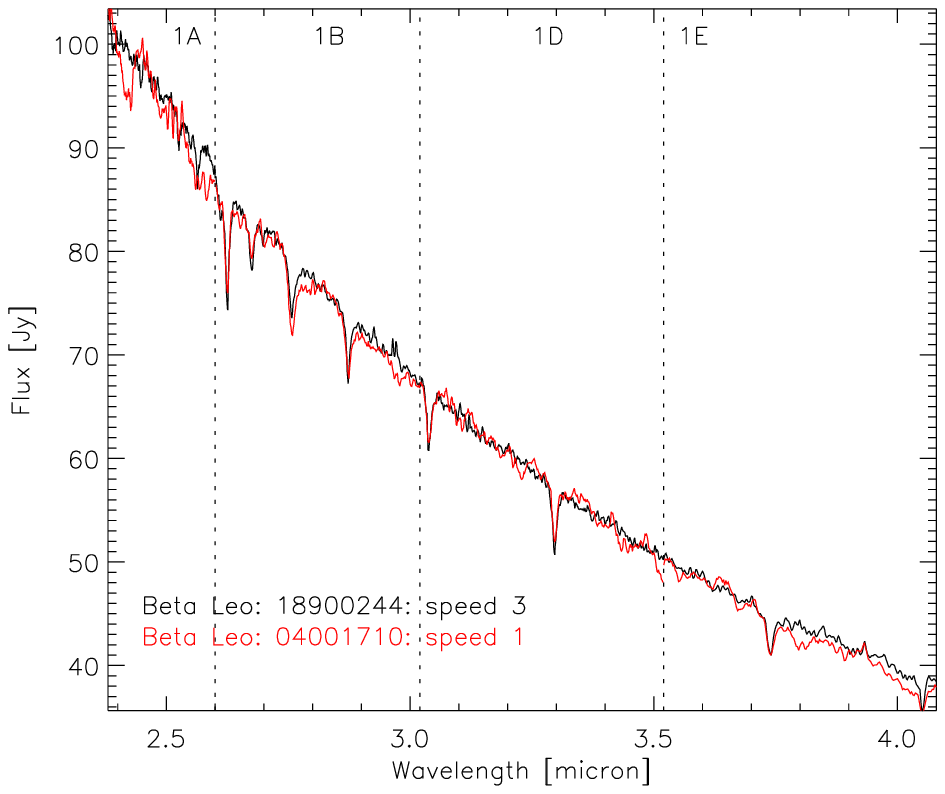}}}
\caption{\label{bleoverscol}Comparison between the AOT01 speed-3
observation of \boldmath {\bf{$\beta$ Leo}} \unboldmath (revolution
189) and the speed-1
observation (revolution 040). The data of the speed-1 observation
have been multiplied by a factor 1.05.}
\end{center}
\end{figure}

\begin{figure}[h!]
\begin{center}
\resizebox{0.5\textwidth}{!}{\rotatebox{90}{\includegraphics{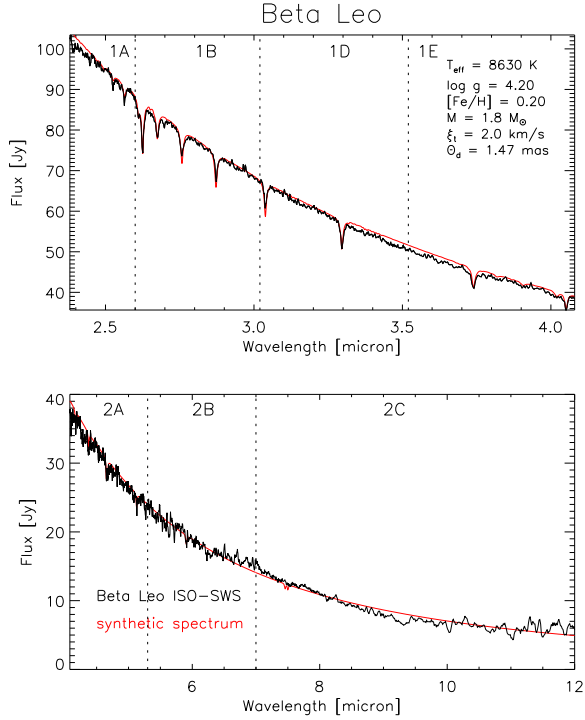}}}
\vspace*{-7ex}
\caption{\label{bleocol} Comparison between band 1 and band 2 of the
ISO-SWS data of \boldmath {\bf{$\beta$ Leo}} \unboldmath (black) and
the synthetic spectrum
(red) with stellar parameters \teff\ = 8630\,K, $\log$ g = 4.20, M
= 1.8\,\Msun, [Fe/H] = 0.20, \vt\ = 2.0\,\kms\ and \ad\ = 1.47\,mas.}
\end{center}
\end{figure}

\begin{figure}[h!]
\begin{center}
\resizebox{0.5\textwidth}{!}{\rotatebox{90}{\includegraphics{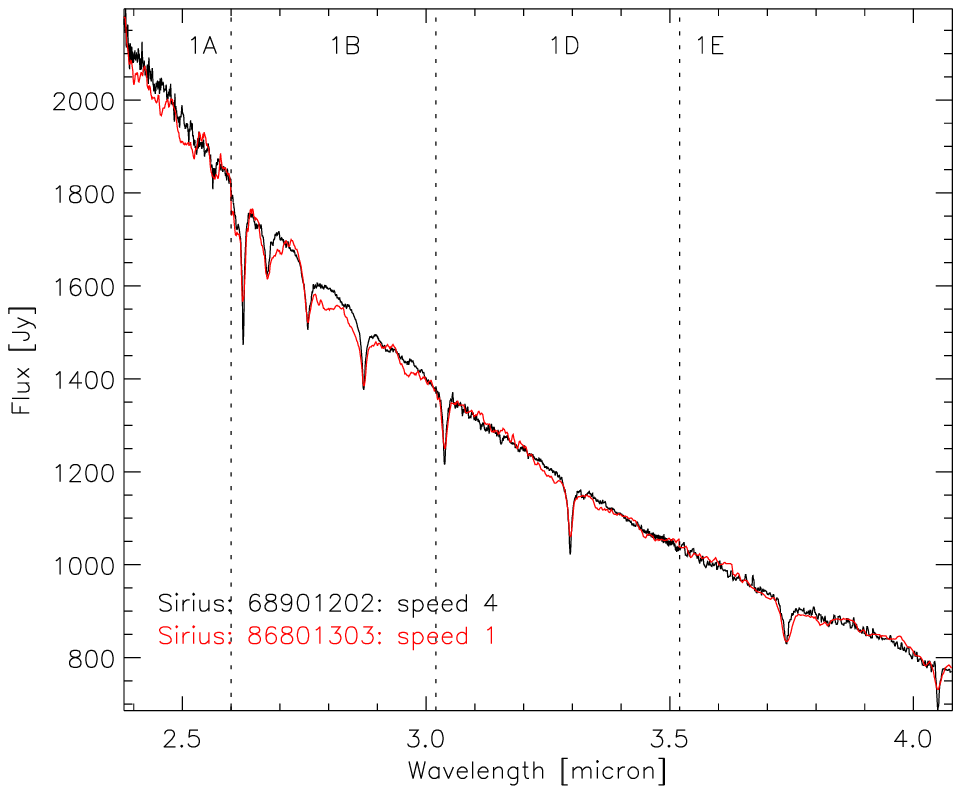}}}
\caption{\label{siriusverscol}Comparison between the AOT01 speed-4
observation of \boldmath {\bf{$\alpha$ CMa}} \unboldmath (revolution
689) and the speed-1
observation (revolution 868). The data of the speed-1 observation
have been multiplied by a factor 1.12.}
\end{center}
\end{figure}

\begin{figure}[h!]
\begin{center}
\resizebox{0.5\textwidth}{!}{\rotatebox{90}{\includegraphics{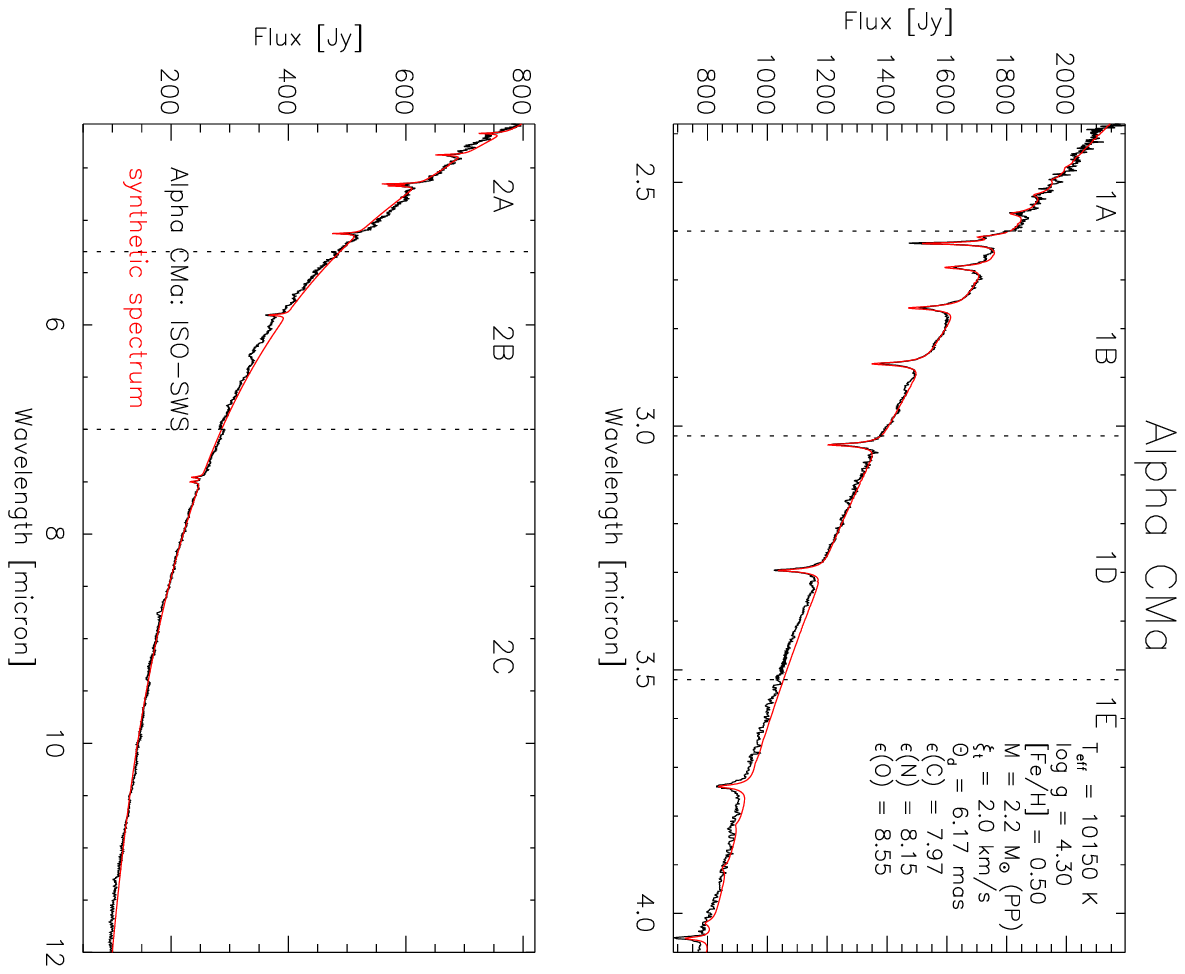}}}
\vspace*{-7ex}
\caption{\label{siriuscol} Comparison between band 1 and band 2 of
the ISO-SWS data of \boldmath {\bf{$\alpha$ CMa}} \unboldmath (black)
and the synthetic
spectrum (red) with stellar parameters \teff\ = 10150\,K, $\log$ g
= 4.30, M = 2.2\,\Msun, [Fe/H] = 0.50, \vt\ = 2.0\,\kms,
$\varepsilon$(C) = 7.97, $\varepsilon$(N) = 8.15, $\varepsilon$(O)
= 8.55 and \ad\ = 6.17\,mas.}
\end{center}
\end{figure}

\begin{figure}[h!]
\begin{center}
\resizebox{0.5\textwidth}{!}{\rotatebox{90}{\includegraphics{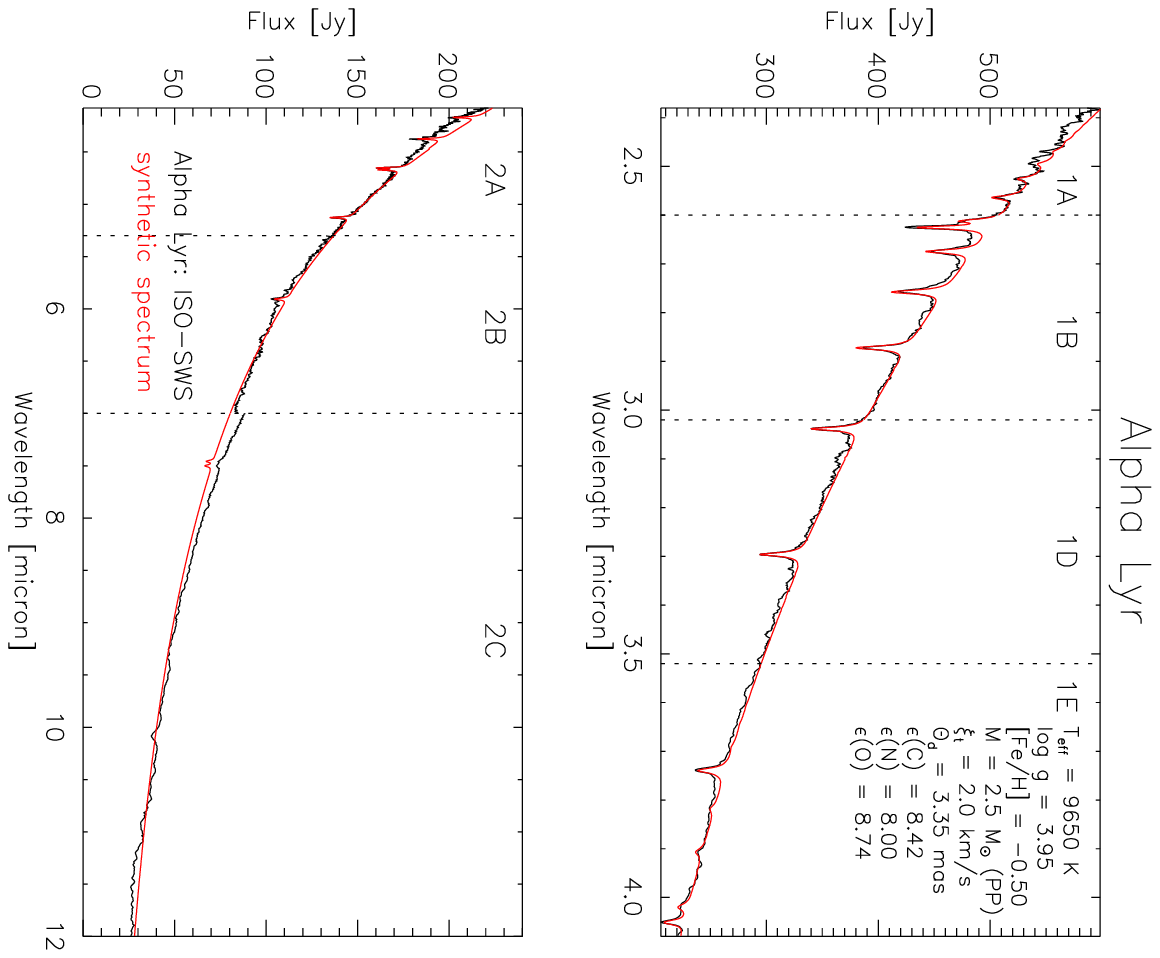}}}
\vspace*{-7ex}
\caption{\label{vegacol} Comparison between band 1 and band 2 of the
ISO-SWS data of \boldmath {\bf{$\alpha$ Lyr}} \unboldmath (black) and
the synthetic spectrum
(red) with stellar parameters \teff\ = 9650\,K, $\log$ g = 3.95, M
= 2.5\,\Msun, [Fe/H] = $-0.50$, \vt\ = 2.0\,\kms,
$\varepsilon$(C) = 8.42, $\varepsilon$(N) = 8.00, $\varepsilon$(O)
=  8.74 and \ad\ = 3.35\,mas.}
\end{center}
\end{figure}

\newpage
\phantom{A}
\newpage
\phantom{A}
\newpage
\phantom{A}
\newpage

\section{Comments on published stellar parameters}\label{literatureapp}

In this appendix, a description of the results obtained by
different authors using various methods is given in chronological order.
One either can look to the quoted reference in the accompanying paper
and than search the description in the chronological (and then
alphabetical) listing below
or one can use the cross-reference table (Table \ref{cross}) to find
all the references for one specific star in this numbered listing.

\begin{table*}
\caption{\label{cross}Cross-reference table in which one can find
all the numbers referring to the papers citing a particular star.}
\begin{center}
\begin{tabular}{l|l} \hline
\rule[-3mm]{0mm}{8mm} name & reference number \\ \hline
\rule[-0mm]{0mm}{5mm}$\alpha$ Cen A &
\ref{Blackwell1977MNRAS.180..177B},
\ref{Flannery1978ApJ...221..175F},
\ref{Kamper1978AJ.....83.1653K},
\ref{England1980MNRAS.191...23E},
\ref{Demarque1986ApJ...300..773D},
\ref{Smith1986ApJ...311..843S},
\ref{Soderblom1986AA...158..273S},
\ref{Abia1988AA...206..100A},
\ref{Edvardsson1988AA...190..148E},
\ref{Volk1989AJ.....98.1918V},
\ref{Furenlid1990ApJ...350..827F},
\ref{McWilliam1990ApJS...74.1075M},
\ref{Chmielewski1992AA...263..219C},
\ref{Engelke1992AJ....104.1248E},
\ref{Pottasch1992AA...264..138P},
\ref{Popper1993ApJ...404L..67P},
\ref{Gadun1994AN....315..413G},
\ref{Neuforge1997AA...328..261N},
\ref{Pourbaix1999AA...344..172P} \\

$\alpha$ Car &
\ref{Brown1974MNRAS.167..121B},
\ref{Code1976ApJ...203..417C},
\ref{Blackwell1977MNRAS.180..177B},
\ref{Linsky1978ApJ...220..619L},
\ref{Luck1979ApJ...232..797L},
\ref{Blackwell1980AA....82..249B},
\ref{Desikachary1982MNRAS.201..707D},
\ref{Boyarchuk1983Ap.....18..228B},
\ref{Lyubimkov1984Ap.....19..385L},
\ref{Luck1985ApJ...298..782L},
\ref{DiBenedetto1987AA...188..114D},
\ref{Russell1989ApJS...70..865R},
\ref{Spite1989AA...210...25S},
\ref{Volk1989AJ.....98.1918V},
\ref{McWilliam1990ApJS...74.1075M},
\ref{Achmad1991AA...249..192A},
\ref{ElEid1994AA...285..915E},
\ref{Gadun1994AN....315..413G},
\ref{Hill1995AA...293..347H},
\ref{Smalley1995AA...293..446S} \\

$\beta$ Leo &
\ref{Brown1974MNRAS.167..121B},
\ref{Code1976ApJ...203..417C},
\ref{Moon1985ApSS.117..261M},
\ref{Moon1985MNRAS.217..305M},
\ref{Lester1986ApJS...61..509L},
\ref{Malagnini1990AAS...85.1015M},
\ref{Napiwotzki1993AA...268..653N},
\ref{Smalley1993AA...271..515S},
\ref{Holweger1995AA...303..819H},
\ref{Smalley1995AA...293..446S},
\ref{Sokolov1995AAS..110..553S},
\ref{Malagnini1997AA...326..736M},
\ref{Gardiner1999AA...347..876G} \\

$\alpha$ CMa &
1.\ \ref{Blackwell1980AA....82..249B},
2.\ \ref{Popper1980ARAA..18..115P},
3.\ \ref{Bell1981ApJ...248.1031B},
4.\ \ref{Lambert1982ApJ...254..663L},
5.\ \ref{Moon1985ApSS.117..261M},
6.\ \ref{Moon1985MNRAS.217..305M},
7.\ \ref{Sadakane1989PASJ...41..279S},
8.\ \ref{Volk1989AJ.....98.1918V},
9.\ \ref{Lemke1990AA...240..331L},
10.\ \ref{Hill1993AA...276..142H},
11.\ \ref{Smalley1995AA...293..446S},
12.\ \ref{Hui-Bon-Hoa1997AA...323..901H},
13.\ \ref{Rentzsch-Holm1997AA...317..178R},
14.\ \ref{DiBenedetto1998AA...339..858D},
15.\ \ref{Qiu2001Apj...548...953Q} \\

\rule[-3mm]{0mm}{3mm}$\alpha$ Lyr &
\ref{Blackwell1980AA....82..249B},
\ref{Dreiling1980ApJ...241..736D},
\ref{Sadakane1981PASJ...33..189S},
\ref{Lambert1982ApJ...254..663L},
\ref{Moon1985ApSS.117..261M},
\ref{Moon1985MNRAS.217..305M},
\ref{Gigas1986AA...165..170G},
\ref{Volk1989AJ.....98.1918V},
\ref{Adelman1990ApJ...348..712A},
\ref{Lemke1990AA...240..331L},
\ref{Venn1990ApJ...363..234V},
\ref{Hill1993AA...276..142H},
\ref{Napiwotzki1993AA...268..653N},
\ref{Smith1993AA...274..335S},
\ref{Castelli1994AA...281..817C},
\ref{Smalley1995AA...293..446S},
\ref{Sokolov1995AAS..110..553S},
\ref{DiBenedetto1998AA...339..858D},
\ref{Ciardi2001ApJ...559.1147C},
\ref{Qiu2001Apj...548...953Q} \\

\hline
\end{tabular}
\end{center}
\end{table*}

\begin{enumerate}
\item{\label{Brown1974MNRAS.167..121B}
\citet{Brown1974MNRAS.167..121B} have used the stellar interferometer at
Narrabri Observatory to measure the apparent angular diameter of
32 stars. The limb-darkening corrections were based on model
atmospheres.}

\item{\label{Code1976ApJ...203..417C}
\citet{Code1976ApJ...203..417C} have derived the effective
temperature from the absolute
flux distribution and the apparent angular diameter
\citep{Brown1974MNRAS.167..121B}. The
absolute flux distribution has been found by combining observations of
ultraviolet flux with ground-based photometry.}

\item{\label{Blackwell1977MNRAS.180..177B}
\citet{Blackwell1977MNRAS.180..177B} have described the Infrared
Flux Method (IRFM)
to determine the stellar angular diameters and effective temperatures from
absolute infrared photometry. For 28 stars (including $\alpha$ Car, $\alpha$
Boo, $\alpha$ CMa, $\alpha$ Lyr, $\beta$ Peg, $\alpha$ Cen A, $\alpha$ Tau and
$\gamma$ Dra) the angular diameters are deduced. Only for the first four stars
the corresponding effective temperatures are computed.}

\item{\label{Flannery1978ApJ...221..175F}
Using the well-known astrometric properties (e.g.\ parallax,
visual orbit) of the binary $\alpha$ Cen,
\citet{Flannery1978ApJ...221..175F} have deduced the mass of the A
and B component of the binary system. The total and individual
masses should be accurate to about 3\,\%, corresponding to the
possible 1\,\% error in parallax, while the uncertainty of the
mass ratio is somewhat smaller, about 2\,\%. The effective
temperature is computed from ($B-V$) and ($V-I$) colour indices.
For the luminosity different broad-band systems (including the
standard $UBV$, the long-wave $RIJKLMN$ and the six-colour
$UViBGRI$) and narrow-band photometric indices were used. By
analysing the temperature sensitive Ca~I transition, composition
dependent stellar evolution models and \teff-colour relationships,
both the temperature and the (enhanced) metallicity are
ascertained.}

\item{\label{Kamper1978AJ.....83.1653K}
\citet{Kamper1978AJ.....83.1653K} have determined the proper motion
and parallax of the $\alpha$ Centauri system by using all
available observations till 1971. The new mass ratio together with
the period and the semi-major axis yielded then the total mass of
the system and so the individual masses of $\alpha$ Cen A and
$\alpha$ Cen B.}

\item{\label{Linsky1978ApJ...220..619L}
\citet{Linsky1978ApJ...220..619L} have estimated the effective
temperatures for the programme stars using the mean of the Johnson
(1966) \teff-($V-I$) transformations, since these are essentially
independent of luminosity.}

\item{\label{Luck1979ApJ...232..797L}
\citet{Luck1979ApJ...232..797L} has performed a chemical analysis
of nine southern supergiant
stars. He obtained spectrograms with the 1.5\,m reflector of Cerro
Tololo Inter-American observatory. Effective temperature, gravity and
microturbulent velocity for the programme stars are determined solely in a
spectroscopic way from the Fe~I and Fe~II lines. The equivalent widths are
calculated through a model atmosphere with these stellar parameters and are then
compared with the observed equivalent widths. The calculation is repeated,
changing the abundance of the species under question, until a match is
achieved.}

\item{\label{Blackwell1980AA....82..249B}
\citet{Blackwell1980A&A....82..249B} have determined the
effective temperature and the
angular diameter for 28 stars using the IRFM method.}

\item{\label{Dreiling1980ApJ...241..736D}
   Several flux-constant, line-blanketed model stellar atmospheres
   have been computed for Vega by
   \citet{Dreiling1980ApJ...241..736D}. The stellar effective
   temperature has been found by comparisons of observed
   and computed absolute fluxes. The Balmer line profiles gave
   the surface gravities, which were consistent with the results from the
   Balmer jump and with the values found from the radius --- deduced from
   the parallax and the limb-darkened angular diameter --- and the (estimated)
   mass. }

\item{\label{England1980MNRAS.191...23E}
The chemical composition of the major components of the bright,
nearby system of $\alpha$ Centauri is derived by
\citet{England1980MNRAS.191...23E} using high-dispersion spectra.
The abundances of 16 elements are found using a differential curve
of growth analysis. Scaled solar LTE model atmospheres for
$\alpha$ Cen A and $\alpha$ Cen B are calculated using effective
temperatures from H$\alpha$ profiles and surface gravities from
line profiles and ionisation equilibria.}

\item{\label{Popper1980ARAA..18..115P}
In his review, \citet{Popper1980ARA&A..18..115P} has discussed
the problems of determining masses from data for eclipsing and visual
binaries. Only individual masses of
considerable accuracy, determined directly from the observational
data, are treated.}

\item{\label{Bell1981ApJ...248.1031B}
   Several flux-constant, line-blanketed model stellar atmospheres
   have been computed for Sirius by
   \citet{Bell1981ApJ...248.1031B}. The stellar effective
   temperature has been found by comparisons of observed
   and computed absolute fluxes. The Balmer line profiles gave
   the surface gravities, which were consistent with the results from the
   Balmer jump and with the values found from the radius --- deduced from
   the parallax and the limb-darkened angular diameter --- and the (estimated)
   mass.}

\item{\label{Sadakane1981PASJ...33..189S}
   \citet{Sadakane1981PASJ...33..189S} derived abundances for
   various elements from observations in the visual and near
   ultraviolet spectra ranges. Their investigation yielded metal
   deficiencies of up to 1.0\,dex; iron was found to be underabundant by
   0.60\,dex, if a solar iron abundance of
   $\varepsilon$(Fe)$ = 7.51$ is assumed.}

\item{\label{Desikachary1982MNRAS.201..707D}
\citet{Desikachary1982MNRAS.201..707D} have taken a weighted mean
of seven determinations from Johnson and Str\"{o}mgren photometry,
absolute spectrophotometry, intensity interferometry and infrared
fluxes. The surface gravity was provided by fits to the H$\gamma$
and H$\delta$-line profiles as well as to Str\"{o}mgren ($b-y$)
and $c_1$ indices. \citet{Luck1985ApJ...298..782L} quoted that
using their parameters (\teff\ = 7500\,K and $\log$ g = 1.5)
reproduces the observed Balmer line profiles about as well as
Desikachary and Hearnshaw's alternative pairing of \teff\ =
7350\,K and $\log$ g = 1.8. Four \'{e}chelle spectrograms were
obtained with a resolution varying between 0.07\,\ang\ and
0.10\,\ang. The microturbulent velocity was measured from 33 Fe~I
lines on the saturated part of the curve of growth ($-5.0 \le
\log(W/\lambda) \le -4.40$). No evidence for depth-dependence of
the microturbulence was found in Canopus. The model atmosphere
grid computed by \citet{Kurucz1979ApJS...40....1K} was used for
this analysis. Hydrostatic and ionisation equilibrium equations
are solved and the line formation problem in LTE is treated to
obtain the equivalent widths of lines of a given element and hence
the abundance.}

\item{\label{Lambert1982ApJ...254..663L}
\citet{Lambert1982ApJ...254..663L} have obtained carbon,
   nitrogen, and oxygen abundances from C~I, N~I and O~I
   high-excitation permitted lines. These results are based on model
   atmospheres and observed spectra. The effective temperature and
   gravity found by \citet{Dreiling1980ApJ...241..736D} and
   \citet{Bell1981ApJ...248.1031B} were used to
   independently determine the metallicity and microturbulent
   velocity.}

\item{\label{Boyarchuk1983Ap.....18..228B}
\citet{Boyarchuk1983Ap.....18..228B} have used published
high-dispersion spectroscopic data in order to analyse Canopus.
The effective temperature and gravity are determined by using the
Balmer lines H$\gamma$ and H$\delta$, the energy distribution in
the continuum and the ionisation equilibrium for V, Cr and Fe.
Using the Fe~I lines yields a microturbulence of 4.5\,\kms, while
a microturbulence of 6.0\,\kms\ was derived from the Ti~II, Fe~II
and Cr~II lines. At that moment, Boyarchuk and Lyubimkov could not
explain this phenomenon, but later on, in 1983, Boyarchuk and
Lyubimkov could explain this as being due to non-LTE effects.}

\item{\label{Lyubimkov1984Ap.....19..385L}
\citet{Lyubimkov1984Ap.....19..385L} have used the model found by
\citet{Boyarchuk1983Ap.....18..228B} to determine the abundances of 21
elements in the atmosphere of
Canopus. The microturbulence derived from the Fe~I lines was used. By comparison
with evolutionary calculations, the mass, radius, luminosity, and age are
found. It is demonstrated that the extension of the atmosphere is small compared
to the radius.}

\item{\label{Luck1985ApJ...298..782L}
\citet{Luck1985ApJ...298..782L} have acquired data for Canopus with
the ESO Coud\'{e} auxiliary telescope and Reticon equipped echelle
spectrograph, with a resolution of 0.05\,\ang. The equivalent
widths were determined by direct integration of the line profiles.
Both photometry and a spectroscopic analysis are used for the
determination of the atmospheric parameters. Broad-($UBVRIJK$) and
narrow-($uvby$) band photometry were available for Canopus. Using
different colour-temperature relations, \teff\ and $\log$ g were
determined in different ways. The uncertainties in these
(photometric) values are estimated to be $\pm 50$\,K in \teff\ and
$\pm 0.2$\,dex in $\log$ g. Spectroscopic estimates of \teff,
$\log$ g and \vt\ are proceeded using the classical requirements:
(1) \teff\ is set by requiring that the individual abundances from
the Fe~I lines are independent of the lower excitation potential;
(2) the requirement that the individual abundances of the Fe~I
lines show no dependence on line strength provides \vt; (3) $\log$
g is determined by forcing the Fe~I and Fe~II lines to give the
same abundance. The formal uncertainties in spectroscopic
parameter determinations are typically $\pm 200$\,K in \teff, $\pm
0.3$\,dex in $\log$ g and $\pm 0.5$\,\kms\ in \vt. Although various
attempts have been performed, the scatter on the iron abundance
determined from the Fe~I lines remained quite high (0.26\,dex). The
photometric values of \teff\ range from 7320 to 7900\,K, with the
spectroscopic value being 7500\,K. From the ($b-y/c_1$) colour, a
surface gravity of 1.80 is ascertained, while the spectroscopic
determination yields a value of 1.50. Finally, the abundances for
C, N, and O have been determined by spectrum synthesis using the
spectroscopic atmospheric values, with the N abundance being
computed from a non-LTE analysis. They quoted that the N-abundance
of \citet{Desikachary1982MNRAS.201..707D} is rather uncertain due to
their LTE analysis and the use of blended or very weak lines.
Although \citet{Luck1985ApJ...298..782L} have performed a non-LTE analysis
to determine to $\varepsilon$(N), they do not have taken
NLTE-effects into account in the determination of the
spectroscopic atmospheric parameters.}

\item{\label{Moon1985ApSS.117..261M}
\citet{Moon1985Ap&SS.117..261M} has found a linear relation
between the visual surface
brightness parameter $F_{\nu}$ and the ($b-y$)$_0$ colour index of $ubvy\beta$
photometry for spectral types later than G0. Using this relation, tables of
intrinsic colours and indices, absolute magnitude and stellar radius are given
for the ZAMS and luminosity classes Ia - V over a wide range of spectral
types.}

\item{\label{Moon1985MNRAS.217..305M}
From empirically calibrated $uvby\beta$ grids,
\citet{Moon1985MNRAS.217..305M} have determined the effective
temperature and
surface gravity of a sample of stars. The authors divided the
temperature range into three region (\teff\ $\le 8500$\,K, \teff\
$\ge 11000$\,K and the region in between these two temperature
values) and have given for every region one grid. Comparison with
fundamental measurements of \teff\ and $\log$ g
\citep{Code1976ApJ...203..417C} shows an excellent agreement, while
Balona's formula \citep{Balona1984MNRAS.211..973B}
for B stars gives values with a mean difference of $110 \pm 360$\,K
for the temperature and $0.22 \pm 0.15$\,dex for the gravity. A
program for the analysis of photometric data based on these grids
has been presented by
\citet{Moon1985MNRAS.217..305M}. \citet{Napiwotzki1993A&A...268..653N}
quoted however that they have noted discrepancies between the
values of \teff\ and $\log$ g derived from the published grid and
the values derived from the polynomial fits used in the program of
\citet{Moon1985MNRAS.217..305M}. Therefore
\citet{Napiwotzki1993A&A...268..653N} have completely rewritten this program.}

\item{\label{Demarque1986ApJ...300..773D}
Using the absolute parallax and the observed apparent
magnitudes, \citet{Demarque1986ApJ...300..773D} have calculated the absolute
magnitudes and masses for the components of $\alpha$ Cen A and
$\alpha$ Cen B.}

\item{\label{Gigas1986AA...165..170G}
Based on high-dispersion spectra covering the wavelength range
   3050 -- 6850 {\AA}, a model-atmosphere analysis of the Fe~I/Fe~II
   spectrum of Vega has been carried out by
   \citet{Gigas1986A&A...165..170G} taking into account departures
   from LTE. The parameters \teff$ = 9500$\,K and $\log {\rm{g}} =
   3.9$ derived by \citet{Lane1984ApJ...281..723L} have been
   adopted. The microturbulence has been derived in the usual way by
   varying \vt\ until all iron lines yielded (almost) the same metal
   abundance independent of the equivalent width. After various test
   calculations it has been found that the best fit is given by a
   microturbulent velocity decreasing with optical depth.}

\item{\label{Lester1986ApJS...61..509L}
\citet{Lester1986ApJS...61..509L} have performed a calibration of the
effective temperature and gravity using the Str\"{o}mgren
$ubvy\beta$ indices based on the line-blanketed LTE stellar
atmospheres from \citet{Kurucz1979ApJS...40....1K}. The indices have
been placed on
the standard systems using the ultraviolet and visual energy
distributions of the secondary spectrophotometric standards. For
these standard stars, the effective temperature and surface
gravity have been determined by finding the model atmosphere which best
matched the observed visual and ultraviolet energy distribution.
They have shown that the common practice of using a single
standard star to effect the transformation of the computed indices
to the standard system produces systematic errors.}

\item{\label{Smith1986ApJ...311..843S}
From a limited set of high-quality data,
\citet{Smith1986ApJ...311..843S} have determined
the physical parameters of $\alpha$ Cen A and $\alpha$ Cen B. Therefore a
conventional analysis based on LTE model atmospheres derived from the
Holweger-M\"{u}ller solar atmosphere has been used. From the data they have
constructed a series of logarithmic abundance against effective temperature
diagrams, each diagram corresponding to fixed values of microturbulence and
surface gravity. The region or `neck' where these lines converge indicates the
values of the effective temperature and the metallicity.}

\item{\label{Soderblom1986AA...158..273S}
\citet{Soderblom1986A&A...158..273S} has compared high-resolution and high
signal-to-noise observations of H$\alpha$ in the Sun, $\alpha$ Cen
A and $\alpha$ Cen B to  models in order to derive
temperatures. These temperatures were then used to calculate the radii of these
stars from the luminosity values given by \citet{Flannery1978ApJ...221..175F}.}

\item{\label{DiBenedetto1987AA...188..114D}
\citet{DiBenedetto1987A&A...188..114D} used Michelson interferometry by
the two-telescope baseline located at CERGA. Combining this
angular diameter with the bolometric flux F$_{\mathrm{bol}}$
(resulting from a directed integration using the trapezoidal rule
over the flux distribution curves, after taking  interstellar
absorption into account) they found the effective temperature,
which was in good agreement with results obtained from the lunar
occultation technique. \citet{DiBenedetto1998A&A...339..858D}
 calibrated the surface
brightness-colour correlation using a set of high-precision
angular diameters measured by modern interferometric techniques.
The stellar sizes predicted by this correlation were then combined
with bolometric-flux measurements, in order to determine
one-dimensional (T, $V-K$) temperature scales of dwarfs and giants.
Both measured and predicted values for the angular diameter are
listed.}

\item{\label{Abia1988AA...206..100A}
\citet{Abia1988A&A...206..100A} have obtained high-resolution, high
signal-to-noise spectra of 23 disk stars. Values for the effective
temperature were derived from photometric indices ($b-y$) and
($R-I$). The mean of these values, given by these two photometric
indices, was used as effective temperature. For $\alpha$ Cen A,
the value given by \citet{Soderblom1986A&A...158..273S} was used. The
$\log$ g values
were derived from the ($b-y$) and $c_1$ indices. A microturbulence
parameter of $1.5$\,\kms\ was adopted for all the stars for which the
value was not found in the literature. Two parallel approaches
were used to determine the abundances: method (a) in which
the equivalent width of lines were fitted to curves of growth derived
from the model atmosphere adopted for each star, and method (b) in
which the synthetic spectra were fitted to the observed lines by
interactive fitting using a high-resolution graphics terminal. In
no case did the two methods disagree by more than $0.03$\,dex in the
derived abundances.}

\item{\label{Edvardsson1988AA...190..148E}
\citet{Edvardsson1988A&A...190..148E}determined logarithmic
surface gravities from the analysis of pressure broadened wings of
strong metal lines. Comparisons with trigonometrically determined
surface gravities give support to the spectroscopic results.
Surface gravities determined from the ionisation equilibria of Fe
and Si are found to be systematically lower than the strong line
gravities, which may be an effect of errors in the model
atmospheres, or departures from LTE in the ionisation equilibria.
When the effective temperature of $\alpha$ Cen A derived by
\citet{Smith1986A&A...165..126S} would be used, instead of the
used 5750\,K, the deduced gravity should only change by an amount
$\le +0.04$\,dex.}

\item{\label{Fracassini1988BICDS..35..121F}
\citet{Fracassini1988BICDS..35..121F} have made a catalogue of stellar
apparent diameters and/or absolute radii, listing 12055 diameters
for 7255 stars. Only the most extreme values are listed.
References and remarks to the different values of the angular
diameter and radius may be found in this catalogue. Also here
these angular diameter values are given in italic mode when
determined from direct methods and in normal mode for indirect
(spectrophotometric) determinations.}

\item{\label{Russell1989ApJS...70..865R}
\citet{Russell1989ApJS...70..865R} have derived initial estimates for
the physical parameters of their programme stars from photometry in
the medium-band\-width $ubvy$ Str\"{o}mgren system. For Canopus
the physical parameters derived by \citet{Boyarchuk1983Ap.....18..228B}
were then used (\teff = 7400\,K, $\log$ g = 1.9) to derive the
abundances from spectroscopic observations made with the 1.88m
telescope in Canberra. Therefore the analysis program WIDTH6 --- a
derivative of Kurucz's ATLAS5 code \citep{Kurucz1970SAOSR.308.....K}, using the
classical assumptions of LTE, hydrostatic equilibrium and a
plane-parallel atmosphere --- was used. The mass was derived from
evolutionary models and the absolute visual magnitude has been
determined from the observed, visual magnitude, the visual
interstellar extinction and the distance modulus. Using the
bolometric corrections, the bolometric magnitude has been
determined, which then yields, in conjunction with the effective
temperature, the radius of the star.}

\item{\label{Sadakane1989PASJ...41..279S}
\citet{Sadakane1989PASJ...41..279S} have analysed the
   high-resolution spectral atlas of Sirius published by
   \citet{Kurucz1979Smith}. Using as effective temperature \teff$ =
   10000$\,K and surface gravity $\log {\rm{g}} = 4.30$ the abundances
   of 19 ions were determined by use of the WIDTH6 program of Kurucz. By
   requiring that the abundance is independent of the equivalent
   width, the microturbulence was obtained.}

\item{\label{Spite1989AA...210...25S}
\citet{Spite1989A&A...210...25S} have used an ESO-CES
spectrograph spectrum of Canopus.
The temperature was adopted from \citet{Luck1985ApJ...298..782L}. The
surface gravity
was determined by forcing the Fe~I and the Fe~II lines to yield the same
abundance. The microturbulence was derived from the Fe~I lines and was assumed
to apply to all species. Using these parameters the metallicity was
ascertained.}

\item{\label{Volk1989AJ.....98.1918V}
\citet{Volk1989AJ.....98.1918V} determined the effective temperature
directly from the literature values of angular-diameter
measurements and total-flux observations (also from literature).
The distance was taken from the Catalog of Nearby Stars
\citep{Gliese1969VeARI..22....1G} or from the Bright Star Catalogue
\citep{Hoffleit1982}. }

\item{\label{Adelman1990ApJ...348..712A}
   An elemental abundance analysis of Vega has been performed by
   \citet{Adelman1990ApJ...348..712A} using high signal-to-noise
   2.4\,{\AA}\,mm$^{-1}$ Reticon observations of the region
   $\lambda\lambda$4313 -- 4809. The effective temperature and
   surface gravity were adopted from
   \citet{Kurucz1979ApJS...40....1K}. The program WIDTH6 of
   \citet{Kurucz1993KurCD..13.....K} was used to deduce the abundances
   of metal lines from the measured equivalent widths and adopted
   model atmospheres. The adopted value for the microturbulence is the
   mean of all the Fe~I and Fe~II values.}

\item{\label{Furenlid1990ApJ...350..827F}
\citet{Furenlid1990ApJ...350..827F} have used high-dispersion
Reticon spectra to perform a differential analysis between the Sun
and $\alpha$ Cen. The model atmosphere analysis was carried out
using the WIDTH6 program by Kurucz. The program was used in an
iterative mode, where abundances, effective temperature, surface
gravity and microturbulence are treated as free parameters, and
the measured equivalent-width values together with the appropriate
atomic constants are the fixed input parameters. Four specific
criteria define a consistent solution: the derived abundances must
be independent of (1) the excitation potential of the lines; (2)
the equivalent-width values of the lines; (3) the optical depth of
formation of the lines; and (4) the level of ionisation of
elements with lines in more than one stage of ionisation.}

\item{\label{Lemke1990AA...240..331L}
   Based on high-resolution Reticon spectra
   \citet{Lemke1990A&A...240..331L} has derived abundances of C, Si,
   Ca, Sr, and Ba for 16 sharp lined, main sequence A
   stars. Str\"{o}mgren photometry was converted into effective
   temperature and gravity by means of the calibration of
   \citet{Moon1985MNRAS.217..305M}. From these parameters, model
   atmospheres were computed with the ATLAS6 code of
   Kurucz. Equivalent widths were measured by direct integration of
   the data. A program computed the quantity $\log{gf\varepsilon}$ in
   such a way that computed and observed equivalent widths
   agree. Optionally, NLTE departure coefficients for calcium and
   barium could be taken into account.}

\item{\label{Malagnini1990AAS...85.1015M}
\citet{Malagnini1990A&AS...85.1015M} have used spectrophotometric
data (in the wavelength range from 3200\,{\AA} to 10000\,{\AA}) and
trigonometric parallaxes to determine the stellar parameters.
Using the \citet{Kurucz1979ApJS...40....1K} models, they have
performed a fitting
procedure which permits to obtain, simultaneously, accurate
estimates not only of the effective temperature and apparent
angular diameter, but also for the $E(B-V)$ excess for \teff\ $>
9000$\,K. From the angular diameter and the parallax, the stellar
radius is computed and so the luminosity is determined. To derive
the mass and the surface gravity, they have compared the stellar
position in the H-R diagram with theoretical evolutionary
tracks. An uncertainty of 0.15\,dex is derived for $\log$ g. By
taking into account contributions from different sources to the
total error, the average uncertainty affecting the stellar
effective temperature, radius, and luminosity is on the order of 2\,\%, 16\,\%
and, 35\,\% respectively.}

\item{\label{McWilliam1990ApJS...74.1075M}
\citet{McWilliam1990ApJS...74.1075M} based his results on
high-resolution spectroscopic observations with resolving power
40000. The effective temperature was determined from empirical and
semi-empirical results found in the literature and from broad-band
Johnson colours. The gravity was ascertained by using the
well-known relation between g, \teff, the mass M and the
luminosity L, where the mass was determined by locating the stars
on theoretical evolutionary tracks. So, the computed gravity is
fairly insensitive to errors in the adopted L. High-excitation
iron lines were used for the metallicity [Fe/H], in order that the
results are less spoiled by non-LTE effects. The author refrained
from determining the gravity in a spectroscopic way (i.e.\ by
requiring that the abundance of neutral and ionised species yields
the same abundance) because {\it{`A gravity adopted by demanding
that neutral and ionised lines give the same abundance, is known
to yield temperatures which are $\sim 200$\,K higher than found by
other methods. This difference is thought to be due to non-LTE
effects in Fe~I lines.'}}. By requiring that the derived iron
abundance, relative to the standard 72 Cyg, were independent of
the equivalent width of the iron lines, the microturbulent
velocity $\xi_t$ was found.}

\item{\label{Venn1990ApJ...363..234V}
\citet{Venn1990ApJ...363..234V}have determined the chemical
   composition of three $\lambda$ Bootis stars and the ``normal'' A
   star Vega. Equivalent widths derived from the spectra --- obtained
   using the Reticon camera at the coud\'{e} focus of the 2.7\,m
   telescope at the McDonald Observatory --- were converted to
   abundances using the program WIDTH6
   \citep{Kurucz1993KurCD..13.....K}. The effective temperature and
   gravity found from a detailed study by
   \citet{Dreiling1980ApJ...241..736D} have been adopted. For the
   microturbulent velocity, they used the value of 2\,\kms\ of
   \citet{Lambert1982ApJ...254..663L}.}

\item{\label{Achmad1991AA...249..192A}
\citet{Achmad1991A&A...249..192A} used several sets of equivalent
widths available in
the literature. More than 2000 lines are then used in a least-square iterative
routine in which \teff, $\log$ g, \vt\ and [Fe/H] are determined
simultaneously. The results are compared with those of other authors. No
significant variation of \vt\ with depth is found.}

\item{\label{Chmielewski1992AA...263..219C}
\citet{Chmielewski1992A&A...263..219C} have used Reticon
spectrograms of $\alpha$ Cen A
and $\alpha$ Cen B to determine their stellar parameters. Like demonstrated by
\citet{Cayrel1985A&A...146..249C} the wings of the hydrogen
H$_{\alpha}$ line profile can be
used to provide the effective temperature relative to the sun. Using the mass
found by \citet{Demarque1986ApJ...300..773D}, the gravity was
calculated from the effective
temperature, the mass, the visual magnitude and the bolometric correction. A
microturbulent velocity parameter of 1\,\kms\ was adopted. Using a curve of
growth the iron and nickel abundances were determined.}

\item{\label{Engelke1992AJ....104.1248E}
\citet{Engelke1992AJ....104.1248E} has derived a two-parameter
analytical expression approximating the long-wavelength (2 --
60\,\mic) infrared continuum of stellar calibration standards.
This generalised result is written in the form of a Planck
function with a brightness temperature that is a function of both
observing wavelength and effective temperature. This function is
then fitted to the best empirical flux data available, providing
thus the effective temperature and the angular diameter.}

\item{\label{Pottasch1992AA...264..138P}
\citet{Pottasch1992A&A...264..138P} reported on the detection of
solar-like p-modes of
oscillation with a period near 5 minutes on $\alpha$ Cen. Using this result, the
radius is derived, which gives, in conjunction with the effective temperature,
the stellar luminosity.}

\item{\label{Hill1993AA...276..142H}
\citet{Hill1993A&A...276..142H} have used spectra obtained with
   the coud\'{e} spectrograph of the Dominion Astrophysical
   Observatory 1.2\,m telescope. The parameters for the model
   atmospheres for the spectrum
   synthesis were chosen using $uvby{\rm{H}}\beta$ photometry. As
   error bars on these atmospheric parameters the values as derived by
   \citet{Lemke1990A&A...240..331L} were taken. The model atmospheres
   were then obtained by interpolating in \teff\ and $\log$ g within
   the grid of plane-parallel, line-blanketed, LTE model atmospheres
   published by \citet{Kurucz1979ApJS...40....1K}. These atmospheres
   assume a depth-independent microturbulence of 2\,\kms\ and a solar
   composition. The spectrum synthesis is performed by a new program,
   which searches for the values of the microturbulence, the radial
   velocity, $v \sin{i}$, and selected abundances by minimising the
   mean square difference between the observed and synthetic
   spectrum.}

\item{\label{Napiwotzki1993AA...268..653N}
\citet{Napiwotzki1993A&A...268..653N} have performed a critical
examination of the determination of stellar temperatures and
surface gravity by means of the Str\"{o}mgren $uvby\beta$
photometric system in the region of main-sequence stars. In
particular, the calibrations of \citet{Moon1985MNRAS.217..305M},
\citet{Lester1986ApJS...61..509L} and
\citet{Balona1984MNRAS.211..973B} are discussed. For the selection
of temperature standards, only those stars were included for which
an integrated-flux temperature was available. The temperatures had
to be based on measurements of the absolute integrated flux which
include both the visual and ultraviolet region. The angular
diameter had to be determined by using the V magnitude or by using
the method proposed by \citet{Malagnini1986A&A...162..140M}, who fitted model
spectra to observed spectra by varying \teff, $\log$ g and \ad.
Results from the IRFM method were excluded due to systematic errors
which seem to destroy the reliability of the results obtained by
the IRFM method. \citet{Napiwotzki1993A&A...268..653N} quoted that the
resulting IRFM temperatures are too low by 1.6 -- 2.8\,\%,
corresponding to angular diameters which are too large by 3.5 -- 5.9\,\%.
In the tables with literature values for $\beta$ Leo and Vega the mean
value of the quoted integrate-flux
temperatures is listed. The photometric temperature values are
then checked against the integrate-flux temperature values. For
the gravity calibration, the authors have determined the surface
gravity by fitting theoretical profiles of hydrogen Balmer lines
\citep{Kurucz1979ApJS...40....1K} to the observations. These
spectroscopic gravities
were then compared with gravities derived from photometric
calibrations. From their results, they recommended the Moon and
Dworetsky calibration, if corrected for gravity deviation. The
final statistical error of the temperature determination ranges
from 2.5\,\% for stars with \teff\ $\le 11000$\,K up to 4\,\% for
\teff\ $\ge 20000$\,K, while the accuracy for the gravity
determinations ranges from $\sim 0.10$\,dex for early A stars to
$\sim 0.25$ dex for hot B stars.}

\item{\label{Popper1993ApJ...404L..67P}
\citet{Popper1993ApJ...404L..67P}
has determined the masses of G -- K main-sequence
stars by observations of detached eclipsing binaries of short
period with the CCD-echelle spectrometer. A mass of 1.14\,\Msun\
was found for $\alpha$ Cen A, which results in a gravity of $\log
{\rm{g}} = 4.3$ when a radius-value found in the literature is used.}

\item{\label{Smalley1993AA...271..515S}
\citet{Smalley1993A&A...271..515S} have presented a detailed
investigation into the methods of determining the atmospheric
parameters of stars in the spectral range A3 -- F5. A comparison is
made between atmospheric parameters derived from Str\"{o}mgren
$uvby\beta$ photometry, from spectrophotometry, and from Balmer
line profiles.  The photometric results found by
\citet{Relyea1978ApJS...37...45R}, \citet{Moon1985MNRAS.217..305M},
\citet{Lester1986ApJS...61..509L}, and \citet{Kurucz1991sabc.conf..441K}
are confronted with each other. All the various photometric
calibrations give generally the same \teff\ and $\log$ g to within
$\pm 200$\,K and $\pm 0.2$\,dex respectively. The model $m_0$ colours (which are
sensitive to the metal abundance) do however not adequately
reproduce the observed values due to inadequacies in the opacities
of the \citet{Kurucz1979ApJS...40....1K} models. Another reason for
poorer results of any of the existing grids of model atmospheres to
reproduce $m_0$ is the fact that this index is also quite sensitive to
the onset of convection which affects any prediction of $m_0$ for
stars later than about A5. Experiments with different
treatments of convection point towards convection as the remaining
major source of uncertainty for the determination of fundamental
parameters from photometric calibrations using model atmospheres in
this temperature region of the HR diagram
\citep{Smalley1997A&A...328..349S}. 

By fitting optical and ultraviolet
spectrophotometric fluxes, the authors have derived the values of
\teff\ and $\log$ g. These spectroscopic values for \teff\ and
$\log$ g, corresponding to two different metallicities, are the
first two values listed for $\beta$ Leo. Using the new
\citet{Kurucz1991sabc.conf..441K}
model fluxes instead of the \citet{Kurucz1979ApJS...40....1K} model fluxes,
yields \teff-values which differ by only $\pm 100$\,K, but the
gravity is higher by typically 0.2 -- 0.3\,dex. The obtained
spectrophotometric values for \teff\ and $\log$ g are in good
agreement with the results from the $uvby\beta$ photometry, but
are systematically lower than the spectrophotometric \teff\ and
$\log$ g given by \citet{Lane1984ApJ...281..723L}. The reason for this
discrepancy was found to be an insufficient allowance for the
metal abundance by \citet{Lane1984ApJ...281..723L}. A third method was
based on medium-resolution spectra of H$\beta$ and H$\gamma$ line
profiles in order to obtain the \teff\ for the 52 A and F stars.
These results for two different metallicities  are the last two
values listed for $\beta$ Leo. These results are in good agreement
with the photometric \teff\ and $\log$ g values. The authors have
concluded that the values of \teff\ and $\log$ g determined from
photometry are extremely reliable and not significantly affected
by the metallicity.}

\item{\label{Smith1993AA...274..335S}
The abundances of five iron-peak elements (chromium through
   nickel) are derived by \citet{Smith1993A&A...274..335S} by
   spectrum-synthesis analysis of co-added high-resolution IUE
   spectra. The effective temperature and gravity were derived from
   calibrations of Str\"{o}mgren and Geneva photometric systems based
   on spectroscopically normal stars and solar-metallicity model
   atmospheres. Further constraints on the effective temperature and
   surface gravity of the programme stars were obtained by fitting the
   predictions of \citet{Kurucz1979ApJS...40....1K} solar-metallicity
   model atmospheres to de-reddened spectrophotometric scans and
   H$\gamma$ profiles of the programme stars taken primarily from the
   literature. The adopted atmospheric parameters are mean values from
   the photometric and `best-fit' spectroscopic analyses. The
   microturbulence parameter was taken from fine analyses of
   visual-region spectra in the literature. Elemental abundances were
   then determined by interactively fitting the observations with LTE
   synthetic spectra computed using the
   \citet{Kurucz1979ApJS...40....1K} models.}

\item{\label{Castelli1994AA...281..817C}
\citet{Castelli1994A&A...281..817C} have compared blanketed LTE
   models for Vega computed both with the opacity distribution
   function method and the opacity-sampling method. The stellar
   parameters (\teff, $\log$ g and [Fe/H]) were fixed by comparing the
   observed ultraviolet, visual, and near
   infrared flux with the computed one and by comparing observed and
   computed Balmer profiles. A microturbulent velocity of 2\,\kms\ was
   assumed. The model parameters for Vega depend on the amount of
   reddening and on the helium abundance.}

\item{\label{ElEid1994AA...285..915E}
The effective temperature, surface gravity and mass of
\citet{Harris1984ApJ...285..674H}
were used by \citet{ElEid1994A&A...285..915E}. He noted a
correlation between the
$^{16}$O/$^{17}$O ratio and the stellar mass and the \cc\ ratio and the stellar
mass for evolved stars. Using this ratio
in conjunction with evolutionary tracks, El Eid has determined the mass.}

\item{\label{Gadun1994AN....315..413G}
\citet{Gadun1994AN....315..413G} has used model parameters and
equivalent widths of Fe~I and Fe~II lines for $\alpha$ Cen,
$\alpha$ Boo, and $\alpha$ Car found in literature. It turned out
that the Fe~I lines were very sensitive to the temperature
structure of the model and that iron was over-ionised relative to
the LTE approximation due to the near-ultraviolet excess
$J_{\nu}-B_{\nu}$. Since the concentration of the Fe~II ions is
significantly higher than the concentration of the neutral iron
atoms, the iron abundance was finally determined using these Fe~II
lines. It is demonstrated that there is a significant difference
in behaviour of \vt\ from the Fe~I lines for solar-type stars,
giants, and supergiants. The microturbulent velocity decreases in
the upper photospheric layers of solar-type stars, in the
photosphere of giants (like Arcturus) \vt\ has the tendency to
increase and in Canopus, a supergiant, a drastic growth of \vt\ is
seen. This is due to the combined effect of convective motions and
waves which form the base of the small-scale velocity field. The
velocity of convective motions decreases in the photospheric
layers of dwarfs and giants, while the velocity of waves increases
due to the decreasing density. In solar-type stars the convective
motion penetrates in the line-forming region, while the behaviour
of \vt\ in Canopus may be explained by the influence of gravity
waves. The characteristics of the microturbulence determined from
the Fe~II lines differ from that found with Fe~I lines. These
results can be explained by 3D numerical modelling of the
convective motions in stellar atmospheres, where it is shown that
the effect of the lower gravity is noticeable in the growth of
horizontal velocities above the stellar `surface' (in the region
of Fe~I line formation). But in the Fe~II line-forming layers the
velocity fields are approximately equal in 3D model atmospheres
with a different surface gravity and same \teff. Both values for
\vt\ derived from the Fe~I and Fe~II lines are listed. If the
microturbulence varies, the values of \vt\ are given going outward
in the photosphere.}


\item{\label{Hill1995AA...293..347H}
\citet{Hill1995A&A...293..347H} have taken the equivalent widths
for Canopus from \citet{Desikachary1982MNRAS.201..707D},
\citet{Luck1985ApJ...298..782L} and
\citet{Spite1989A&A...210...25S}. As a first approach, the
effective temperature was estimated from a ($B-V$)-\teff\
calibration. This first guess for the temperature was checked,
when possible, by fitting the observed and computed profiles of
the H$\alpha$ wings. The final \teff-values were determined by
requiring the Fe~I abundances to be independent of the excitation
potential of the lines. The formal uncertainty in this
determination is about $\pm 200$\,K. The surface gravity was
adjusted to obtain the same iron abundance from weak Fe~I and
Fe~II lines. This gives a maximum uncertainty of 0.3\,dex on the
gravity. The microturbulent velocity was determined by requiring
the abundances derived from the Fe~I lines to be independent of
the line's equivalent width. The uncertainty on \vt\ is of the
order of 0.5\,\kms. By varying these three stellar
parameters, it was seen that no dramatic changes appear upon
gravity and microturbulence variation. Significant changes in
[Fe/H] values take place as a result of temperature variation, but
the relative elemental abundances are changed negligibly. The
error in abundances is of the order of $\pm 0.1$\,dex to $\pm
0.2$\,dex, but the error in abundances relative to iron only
ranges from $\pm 0.01$\,dex to $\pm 0.1$\,dex, depending on the
element. The errors listed in Table 3 for $\alpha$ Car are the
{\it{intrinsic}} errors. Taking into account the overionisation
due to NLTE-effects, yields for a star with similar atmospheric
parameters the same temperature, but an increase in gravity and
\vt. Model atmospheres were interpolated in a grid using the
MARCS-code of \citet{Gustafsson1975A&A....42..407G}. }

\item{\label{Holweger1995AA...303..819H}
\citet{Holweger1995A&A...303..819H} have included $\beta$ Leo in their
sample
of normal main-sequence B9.5 -- A6 stars whose infrared excess indicates the
presence of a circumstellar dust disk. Using published Str\"{o}mgren photometry
and the calibration of \citet{Napiwotzki1993A&A...268..653N} the
temperature and surface gravity was determined. The error-bars are the
ones quoted by \citet{Napiwotzki1993A&A...268..653N}.}

\item{\label{Smalley1995AA...293..446S}
\citet{Smalley1995A&A...293..446S} have presented an investigation into the
determination of fundamental values of \teff\ and $\log$ g. Using angular
diameters of \citet{Brown1974MNRAS.167..121B} and ultraviolet and
optical fluxes, the
effective temperature was derived.  For stars in eclipsing and visual
binary systems, fundamental values for the gravity were listed.
For stars with \teff\ $> 8500$\,K, fits were
also made to the H$\beta$ profiles to determine $\log$ g. The
Str\"{o}mgren-Crawford $uvby\beta$ photometric system provided a quick and
efficient mean of estimating the atmospheric parameters of B, A, and F
stars. Therefore several model atmosphere calibrations were available.}

\item{\label{Sokolov1995AAS..110..553S}
\citet{Sokolov1995A&AS..110..553S} has used the slope of the
Balmer continuum
between 3200\,{\AA} and 3600\,{\AA} in order to determine the
effective temperature of B, A, and F main-sequence stars.
Therefore, stars were selected from two catalogues of
low-resolution spectra, observed in the wavelength region from
3100\,{\AA} to 7400\,{\AA} with a step of 25\,{\AA}. Based on a
selection of temperature standards found in different literature
sources, the author has determined the relationship between \teff\
and the slope of the Balmer continuum. The effective temperatures
determined in that way are in good agreement with results found by
other authors using different methods. The statistical errors of
the temperature determination range from 4\,\% for stars with
\teff\ $\le 10000$\,K up to 10\,\% for stars with \teff\ $\ge 20000$\,K.}

\item{\label{Hui-Bon-Hoa1997AA...323..901H}
\citet{Hui-Bon-Hoa1997A&A...323..901H} have investigated 11 A
   stars in young open clusters and three field stars by means of
   high-resolution spectroscopy. The effective temperature and surface
   gravity were determined by using the $uvby\beta$ photometry and the
   grids of \citet{Moon1985MNRAS.217..305M}. The model atmosphere is
   then interpolated in the grids of Kurucz' ATLAS9 models
   \citep{Kurucz1993KurCD..13.....K}. The microturbulent velocity
   is obtained by the constraint that all the lines of a same element
   should yield the same abundance.}

\item{\label{Malagnini1997AA...326..736M}
\citet{Malagnini1997A&A...326..736M} have discussed the
uncertainties affecting the determinations of effective
temperature and apparent angular diameters based on the flux
fitting method. Therefore they have analysed the influence of the
indetermination of some fixed secondary parameters (i.e.\ surface
gravity, overall metallicity and interstellar reddening) on the
estimates of the fitted parameters \teff\ and \ad. A database of
visual spectrophotometric data together with a grid of theoretical
models of \citet{Kurucz1993KurCD..13.....K} is used. By varying
the fixed parameters, being $\log$ g, [M/H] and $E(B-V)$, by $\pm
0.5$, $\pm 0.5$\,dex and $\pm 0.02$\,mag respectively, the
uncertainties in the determinations of \teff\ and \ad\ are found
to be of the order of 2\,\% (median values) spanning ranges 0.6 --
5.3\,\% and 1 -- 9\,\% respectively. The authors concluded that
these uncertainties must be taken into account by those scientists
who use the effective temperatures, based on the flux fitting
method, in their analysis of high-resolution spectra in order to
avoid systematic errors in their results on chemical abundances of
individual elements.}

\item{\label{Neuforge1997AA...328..261N}
\citet{Neuforge1997A&A...328..261N} have performed a detailed
spectroscopic analysis of the two components of the binary system
$\alpha$ Centauri on the basis of high-resolution and high
signal-to-noise spectra. The temperatures of the stars have been
determined from the Fe~I excitation equilibrium and checked from
the H$\alpha$ line wings. In each star, the microturbulent
velocity, \vt, is determined so that the abundances derived from
the Fe~I lines are independent of their equivalent widths and the
surface gravity is ascertained by forcing the Fe~II lines to
indicate the same abundance as the Fe~I lines. The abundances are
adjusted until the calculated equivalent width of a line is equal
to the observed one.}

\item{\label{Rentzsch-Holm1997AA...317..178R}
\citet{Rentzsch-Holm1997A&A...317..178R} has determined nitrogen
   and sulphur abundances in 15 sharp-lined `normal' main sequence A
   stars from high-resolution spectra obtained with the Coud\'{e}
   spectrograph (CES) of the ESO 1.4\,m telescope. Stellar parameters
   were adopted from \citet{Lemke1990A&A...240..331L}. Using the
   ATLAS9 model atmospheres of \citet{Kurucz1993KurCD..13.....K},
   detailed non-LTE calculations are performed for each model
   atmosphere.}

\item{\label{DiBenedetto1998AA...339..858D}
\citet{DiBenedetto1998A&A...339..858D} calibrated the surface
brightness-colour correlation using a set of high-precision
angular diameters measured by modern interferometric techniques.
The stellar sizes predicted by this correlation were then combined
with bolometric-flux measurements, in order to determine
one-dimensional ($T$, $V-K$) temperature scales of dwarfs and giants.
Both measured and predicted values for the angular diameter are
listed. }

\item{\label{Gardiner1999AA...347..876G}
Using Str{\o}mgren photometry,
\citet{Gardiner1999A&A...347..876G} estimated $\beta$ Leo to be
slightly overabundant, $ {\rm{[Fe/H]}} = +0.2$\,dex. From their
research, one may conclude that the determination
of the effective temperature and gravity for $\beta$ Leo from Balmer
line profiles \citep[as done by][]{Smalley1995A&A...293..446S} seems to
be rather uncertain, since $\beta$ Leo is located close to the
maximum of the Balmer line width as a function of the effective
temperature and the Balmer lines are very sensitive to $\log {\rm{g}}$
in this region. }

\item{\label{Pourbaix1999AA...344..172P}
Instead of disjoint determinations of the visual orbit, the
mass ratio and the parallax, \citet{Pourbaix1999A&A...344..172P} have
undertaken a simultaneous adjustment of all visual and
spectroscopic observations for $\alpha$ Cen. This yielded for the
first time an agreement between the astrometric and spectroscopic
mass ratio. The orbital parallax differs from all previous
estimates, the Hipparcos one being the closest to their value.}

\item{\label{Ciardi2001ApJ...559.1147C}
Near-infrared (2.2\,\mic) long baseline interferometric
observations of Vega are presented by \citet{Ciardi2001ApJ...559.1147C}. The
stellar disk of the star has been resolved and the limb-darkened
stellar diameter and the effective temperature are derived. The
derived value for the angular diameter agrees well with the value
determined by \citet{Brown1974MNRAS.167..121B}, \ad$ = 3.24 \pm 0.07$\,mas.}

\item{\label{Qiu2001Apj...548...953Q}
On the basis of high-resolution echelle spectra obtained with the
   Coud\'{e} Echelle Spectrograph attached to the 2.16\,m telescope at
   the Beijing Astronomical Observatory
   \citet{Qiu2001Apj...548...953Q} have performed an elemental
   abundance analysis of Sirius and Vega. For the effective temperature,
   they adopted the values obtained by \citet{Moon1985MNRAS.217..305M}
   based on $uvby\beta$ photometry. The empirical method to determine
   the surface gravity by requiring that Fe~I and Fe~II give the same
   abundance has been employed. The microturbulence has been derived
   in the usual way by requiring that all the iron lines yield the
   same abundance independent of the line strength. Usually, they have
   taken the initial model metallicity from previous published
   analyses. They then have iterated the whole procedure to give the
   convergence value to be the model overall metallicity. Model
   atmospheres generated by te ATLAS9 code
   \citep{Kurucz1993KurCD..13.....K} were used in the abundance
   analysis. The resulting abundance pattern was then compared with
   other published values.}

\end{enumerate}
\aareferences
\end{document}


\newcommand{\ang}{$\rm \AA$}
\newcommand{\tauross}{$\tau_{\mathrm{ross}}$}
\newcommand{\Msun}{M$_{\odot}$}
\newcommand{\Rsun}{R$_{\odot}$}
\newcommand{\Lsun}{L$_{\odot}$}
\newcommand{\be}{\begin{equation}}
\newcommand{\ee}{\end{equation}}
\newcommand{\bee}{\begin{eqnarray}}
\newcommand{\ad}{$\theta_d$}
\newcommand{\vt}{$\xi_t$}
\newcommand{\cc}{$\mathrm{^{12}C/^{13}C}$}
\newcommand{\kn}{$\kappa_{\nu}$}
\newcommand{\lnu}{$l_{\nu}$}
\newcommand{\ha}{H$_{\alpha}$}
\newcommand{\ea}{et al.}
\newcommand{\ene}{\end{eqnarray}}
\newcommand{\teff}{T$_{\mathrm{eff}}$}
\newcommand{\mic}{$\mu$m}
\newcommand{\hoogte}[1]{\rule{0pt}{#1}}
\newcommand{\hminff}{H$^-_{\rm{ff}}$}
\hyphenation{Kol-mo-go-rov}
\hyphenation{Kol-mo-go-rov--Smir-nov}

\def\offinterlineskip{\baselineskip=-1000pt \lineskip=1pt
\lineskiplimit=\maxdimen}
\def\pra{\mathrel{\mathchoice {\vcenter{\offinterlineskip\halign{\hfil$\displaystyle##$\hfil\cr\propto\cr\sim\cr}}}
{\vcenter{\offinterlineskip\halign{\hfil$\textstyle##$\hfil\cr\propto\cr\sim\cr}}}
{\vcenter{\offinterlineskip\halign{\hfil$\scriptstyle##$\hfil\cr\propto\cr\sim\cr}}}
{\vcenter{\offinterlineskip\halign{\hfil$\scriptscriptstyle##$\hfil\cr\propto\cr\sim\cr}}}}}

\def\ga{\mathrel{\mathchoice {\vcenter{\offinterlineskip\halign{\hfil$\displaystyle##$\hfil\cr>\cr\sim\cr}}}
{\vcenter{\offinterlineskip\halign{\hfil$\textstyle##$\hfil\cr>\cr\sim\cr}}}
{\vcenter{\offinterlineskip\halign{\hfil$\scriptstyle##$\hfil\cr
>\cr\sim\cr}}}
{\vcenter{\offinterlineskip\halign{\hfil$\scriptscriptstyle##$\hfil\cr>\cr\sim\cr}}}}}

\def\la{\mathrel{\mathchoice {\vcenter{\offinterlineskip\halign{\hfil$\displaystyle##$\hfil\cr<\cr\sim\cr}}}
{\vcenter{\offinterlineskip\halign{\hfil$\textstyle##$\hfil\cr<\cr\sim\cr}}}
{\vcenter{\offinterlineskip\halign{\hfil$\scriptstyle##$\hfil\cr
<\cr\sim\cr}}}
{\vcenter{\offinterlineskip\halign{\hfil$\scriptscriptstyle##$\hfil\cr><cr\sim\cr}}}}}

\title{ISO-SWS calibration and the accurate modelling of cool-star
atmospheres \thanks{Based on observations with ISO, an ESA project with
instruments funded by ESA Member States (especially the PI countries France,
Germany, the Netherlands and the United Kingdom) and with the participation of
ISAS and NASA.}}
\subtitle{III.\ A0 to G2 stars}

\author{L.~Decin\inst{1}\thanks{\emph{Postdoctoral Fellow of the Fund for
Scientific Research, Flanders}}  \and
B. Vandenbussche\inst{1} \and
C.~Waelkens\inst{1}
K.~Eriksson\inst{2} \and
B.~Gustafsson\inst{2} \and
B.~Plez\inst{3} \and
A.J.~Sauval\inst{4}
}

\offprints{L.\ Decin, e-mail: Leen.Decin@ster.kuleuven.ac.be}

\institute{Instituut voor Sterrenkunde, KULeuven, Celestijnenlaan 200B, B-3001
    Leuven, Belgium
\and
    Institute for Astronomy and Space Physics, Box 515, S-75120 Uppsala, Sweden
\and
    GRAAL - CC72, Universit\'{e} de Montpellier II, F-34095 Montpellier Cedex 5,
France
\and
    Observatoire Royal de Belgique, Avenue Circulaire 3, B-1180 Bruxelles,
    Belgium
}

\date{Received data; accepted date}

\abstract{
Vega, Sirius, $\beta$ Leo, $\alpha$ Car and $\alpha$ Cen A
belong to a  sample of twenty stellar sources used for the
calibration of the detectors of the Short-Wavelength Spectrometer on
board the Infrared Space Observatory (ISO-SWS). While general
problems with the calibration and with the theoretical modelling of
these stars are reported in \citet{Decin2000b}, each of these stars
is discussed individually in this paper. As demonstrated in
\citet{Decin2000b}, it is not possible to deduce the effective
temperature, the gravity and the chemical composition from the ISO-SWS
spectra of these stars. But since ISO-SWS is absolutely calibrated,
the angular diameter (\ad) of these stellar sources can be deduced from
their ISO-SWS spectra, which consequently yields the stellar radius (R),
the gravity-inferred mass (M$_g$) and the luminosity (L) for these
stars. For Vega, we obtained \ad$ = 3.35 \pm 0.20$\,mas, R$ = 2.79 \pm
0.17$\,\Rsun, M$_g = 2.54 \pm 1.21$\,\Msun\ and L$ = 61 \pm 9$\,\Lsun;
for Sirius \ad$ = 6.17 \pm 0.38$\,mas, R$ = 1.75 \pm
0.11$\,\Rsun, M$_g = 2.22 \pm 1.06$\,\Msun\ and L$ = 29 \pm 6$\,\Lsun;
for $\beta$ Leo \ad$ = 1.47 \pm 0.09$\,mas, R$ = 1.75 \pm
0.11$\,\Rsun, M$_g = 1.78 \pm 0.46$\,\Msun\ and L$ = 15 \pm 2$\,\Lsun;
for $\alpha$ Car \ad$ = 7.22 \pm 0.42$\,mas, R$ = 74.39 \pm
5.76$\,\Rsun, M$_g = 12.80^{+24.95}_{-6.35}$\,\Msun\ and L$ = 14573 \pm
2268$\,\Lsun\ and for $\alpha$ Cen A \ad$ = 8.80 \pm 0.51$\,mas, R$ = 1.27 \pm
0.08$\,\Rsun, M$_g = 1.35 \pm 0.22$\,\Msun\ and L$ = 1.7 \pm 0.2$\,\Lsun.
These deduced parameters are confronted with other published values and the
goodness-of-fit between observed ISO-SWS data and the
corresponding synthetic spectrum is discussed.
}

\maketitle
\markboth{L.\ Decin et al.: ISO-SWS and modelling of cool stars}{L.\ Decin et
al.: ISO-SWS and modelling of cool stars}

\defcitealias{Decin2000A&A...364..137D}{Paper~I}
\defcitealias{Decin2000b}{Paper~II}

\section{Introduction}\label{introduction}
In the first two papers of this series \citep[][ hereafter referred
to as Paper~I and Paper~II respectively]{Decin2000A&A...364..137D,
Decin2000b}, a method is described to analyze a sample of ISO-SWS
spectra of standard stars in a consistent way. We did not only
concentrate on the possibility to extract reliable stellar
parameters from the medium-resolution ISO-SWS spectra, but have
also demonstrated where problems in the computation of synthetic
spectra --- based on the MARCS and Turbospectrum code
\citep{Gustafsson1975A&A....42..407G, Plez1992A&A...256..551P,
Plez1993ApJ...418..812P}, version May 1998 --- and in the
calibration of the ISO-SWS detectors destroy the goodness-of-fit
between observed and synthetic spectra (Paper~II). These general
results were based on a sample of 5 warm (\teff $>$
T$_{\mathrm{eff},\odot}$) and 11 cool stars. In this paper, we
will further analyse these 5 warm stars --- $\alpha$ Cen A,
$\beta$ Leo, $\alpha$ Car, Sirius and Vega ---  in order to
extract relevant astrophysical data.

After a description of the general problems for these warm stars in
Sect.\ \ref{sumprob} (as described in Paper~II), we will outline the
method of analysis to deduce different stellar parameters in Sect.\
\ref{stelparameters} (based on the results of Paper~I and
Paper~II). In the different subsections of Sect.\
\ref{stelparameters}, each star will be discussed individually. In
order to assess the observed accuracy, some specific calibration
details will be given. If available, different AOT01
observations{\footnote{Each observation is determined uniquely by its
observation number (8 digits), in which the first three digits represent the
revolution number. The observing data can be calculated from the
revolution number which is the number of days after 17 November
1995.}}  (i.e.\ a full SWS scan at reduced spectral resolution, with
four possible scan speeds) are compared with each other to demonstrate
the calibration precision of ISO-SWS. With these remarks in mind, the
synthetic spectrum based on assumed and deduced parameters is
confronted with the ISO-SWS spectrum. Furthermore, we will discuss why
we have assumed certain parameters and we will confront the deduced
parameters from the ISO-SWS spectra with other literature values.

The appendix of this article is published electronically. Most of the
grey-scale plots in the article are printed in colour in the appendix,
in order to better distuingish the different spectra.


\section{Summary of general discrepancies (Paper~II)}\label{sumprob}

For the warm stars in our sample, the origin of the general
discrepancies between the ISO-SWS and synthetic
spectra could be reduced to 1.\ inaccurate atomic oscillator strengths in
the infrared, 2.\ problematic computation of hydrogenic line broadening, 3.\
fringes at the end of band 1D (3.02 -- 3.52\,\mic), 4.\ inaccurate Relative
Spectral Response Function (RSRF) at the beginning of band 1A (2.38 --
2.60\,\mic) and 5.\ memory effects in band 2 (4.08 -- 12.00\,\mic).

\section{Stellar parameters}\label{stelparameters}

In Paper~I of this series, a method was described to determine
stellar parameters from the band-1 data (2.38 -- 4.08\,\mic) of
ISO-SWS spectra. This method was based on the presence of different
molecular absorbers in this wavelength range, each having their own
characteristic absorption pattern. Since the infrared absorption
pattern of these A0 -- G2 stars is completely dominated by atoms (with the
exception of $\alpha$ Cen A, for which the CO first overtone and
fundamental bands are weakly visible) this method of analysis could
not be applied to these stars. Moreover, it was demonstrated in
Paper~II that there are still quite some problems with the oscillator
strengths of infrared atomic transitions. It was therefore impossible
to determine the effective temperature (\teff), the gravity ($\log$
g), the microturbulence (\vt), the metallicity ([Fe/H]) and the
abundance of carbon, nitrogen and oxygen for these warm stars from
their ISO-SWS spectra. In order to further analyze these spectra, we
have performed a detailed literature study to find accurate values for
these stellar parameters. Using these parameter values, synthetic
spectra were computed for each
target. From the absolutely calibrated ISO-SWS spectra, we then could
deduce the angular diameter (\ad). The angular diameter together with
the Hipparcos'
parallax (with the only exception of $\alpha$ Cen A for which a
more precise parallax  by \citet{Pourbaix1999A&A...344..172P} is
available) then yielded the stellar radius. Together with the assumed
gravity and effective temperature, the gravity-inferred mass (M$_g$)
and the stellar luminosity (L) are derived.

The resultant stellar parameters are summarised in Table
\ref{parameters}. The objects have been sorted by spectral type.
Since the error bars of certain assumed stellar parameters were
necessary for the propagation to the mean error of other deduced
parameters (see Eq.\ (18) in Paper~I), the error bars on all
stellar parameters are given. The mean error on the angular
diameter is estimated from the intrinsic error, the absolute flux
error (10\,\%) and the error in the assumed effective temperature
(see Paper~II). Whenever $\sigma(\log \rm{g}) \ge 0.40$, the lower
and upper limit of the gravity-inferred mass M$_g$ are estimated
as being 2/3 of the {\it maximum} error. In the subsequent
subsections, each star will be discussed individually. A short
description of the methods and/or data used and on the parameters
assumed and deduced by the different authors quoted in next
sections, can be retrieved from
http://\-www.\-ster.\\kuleuven.\-ac.\-be/\-\~\,$\!$leen/\-artikels/\-ISO3/\-appendix.ps.

\begin{table*}[htb]
\caption{\label{parameters} Fundamental stellar parameters
for the selected stars in the sample. The effective temperature
\teff\ is given in K, the logarithm of the gravity in c.g.s.
units, the microturbulent velocity \vt\ in \kms, the angular
diameter in mas, the parallax $\pi$ in mas, the distance D in
parsec, the radius R in \Rsun, the gravity-inferred mass M$_g$ in
\Msun\ and the luminosity L in \Lsun.}
\begin{center}
\leavevmode
\footnotesize
\setlength{\tabcolsep}{1.2mm}
\begin{tabular}{lcccccc}  \hline
\rule[-3mm]{0mm}{8mm}  & \object{$\alpha$ Lyr} & \object{$\alpha$ CMa} &
\object{$\beta$ Leo} & \object{$\alpha$ Car} & \object{$\alpha$ Cen A} \\
\hline \hline
\rule[0mm]{0mm}{5mm}Sp. Type & A0~V & A1~V & A3~Vv & F0~II
& G2~V \\
\teff & $9650 \pm 200$ & $10150 \pm 400$ & $8630
\pm 200$ & $7350 \pm 300$ & $5830 \pm 30$  \\
$\log$ g & $3.95 \pm 0.20$ & $4.30 \pm 0.20$ & $4.20 \pm 0.10$ & $1.80 \pm
0.50$ & $4.35 \pm 0.05$ \\
\vt & $2.0 \pm 0.5$ & $2.0 \pm 0.5$ & $2.0$ & $3.25 \pm 0.25$ & $1.0 \pm 0.1$ \\
${\mathrm{[Fe/H]}}$ & $-0.40 \pm 0.30$ & $0.50 \pm 0.30$ & $0.20$ &
$-0.24 \pm 0.04$ & $0.25 \pm 0.02$ \\
$\varepsilon$(C) & $8.42 \pm 0.15$& $7.97 \pm 0.15$
& $8.76$ & $8.41 \pm 0.10$ & $8.74 \pm 0.05$ \\
$\varepsilon$(N) & $8.00 \pm 0.15$ & $8.15 \pm 0.15$ & $8.25$ & $8.68
\pm 0.05$ & $8.26 \pm 0.09$ \\
$\varepsilon$(O) & $8.74 \pm 0.15$  & $8.55 \pm 0.12$ & $9.13$ & $8.91
\pm 0.10$ & $9.13 \pm 0.06$ \\
\ad & $3.35 \pm 0.20$ & $6.17 \pm 0.38$ & $1.47 \pm 0.09$
& $7.22 \pm 0.42$ & $8.80 \pm 0.51$ \\
$\pi$ & $128.93 \pm 0.55$ & $379.21 \pm 1.58$ & $90.16 \pm 0.89$ & $10.43 \pm
0.53$ & $737.0 \pm 2.6$ \\
D & $7.76 \pm 0.03$ & $2.63 \pm 0.01$ & $11.09 \pm 0.11$ & $95.88 \pm 4.87$ &
$1.36 \pm 0.01$  \\
R & $2.79 \pm 0.17$ & $1.75 \pm 0.11$ & $1.75 \pm 0.11$ & $74.39 \pm
5.76$ & $1.27 \pm 0.08$ \\
M$_g$ & $2.54 \pm 1.21$ & $2.22 \pm 1.06$  & $1.78 \pm 0.46$  &
$12.77^{+24.95}_{-6.35}$ & $1.35 \pm 0.22$ \\
\rule[-3mm]{0mm}{3mm}L & $61 \pm 9$ & $29
\pm 6$ & $15\pm 2$ & $14573 \pm 2268$ & $1.7 \pm 0.2$ \\ \hline
\end{tabular}
\end{center}
\end{table*}

\subsection{${\alpha}$ Cen A{\rm: AOT01, speed 4, revolution 607}}\label{disacen}

\subsubsection{Some specific calibration details}

Since $\alpha$ Cen A is component of a binary, one has to
check the flux contribution of the second component (HIC 71681,
K1~V). From the coordinates of the system in 1997, its proper
motion and the correction for the orbit, one obtains
\[ \left\{ \begin{array}{r@{\quad=\quad}l}
\alpha_A & 14 \mathrm{h}\ 39\mathrm{m}\ 24.13\mathrm{s}\\ \delta_A
& -60 \mathrm{deg}\ 49'\ 17.9''
\end{array} \right. \]
\[ \left\{ \begin{array}{r@{\quad=\quad}l}
\alpha_B & 14\mathrm{h}\ 39\mathrm{m}\ 22.73\mathrm{s} \\ \delta_B
& -60\mathrm{deg}\ 49'\ 30.4''
\end{array} \right. \]
This results in a difference in spacecraft coordinates of dy =
$-12.1241''$ and dz = $10.9342''$. Taking the pointing jitter into
account ($\le 1.5''$), and the fact that the average difference
between  the G and K star in the wavelength range from 1.6\,\mic\
to 11.2\,\mic\ is $0.91^m$ \citep{Cohen1996AJ....112..241C}, one can calculate
that the maximal flux contribution from the K dwarf around 3\,\mic\
is 6~Jy, which is negligible.

The factors, by which the data of the different sub-bands are
multiplied (see Table 3 in \citetalias{Decin2000b}) show a good agreement with
the band-border ratios determined by \citet{Feuchtgruber1998}. The only
exception is band 2C, but this is not so significant, due to the
large scatter for the band-border ratio between band 2B and band
2C \citep[Fig.\ 6 in ][]{Feuchtgruber1998} and the memory effects in band 2.

\subsubsection{Comparison with other AOT01 observations}\label{compobsacena}

\begin{figure}[h!]
\begin{center}
\resizebox{0.5\textwidth}{!}{\rotatebox{90}{\includegraphics{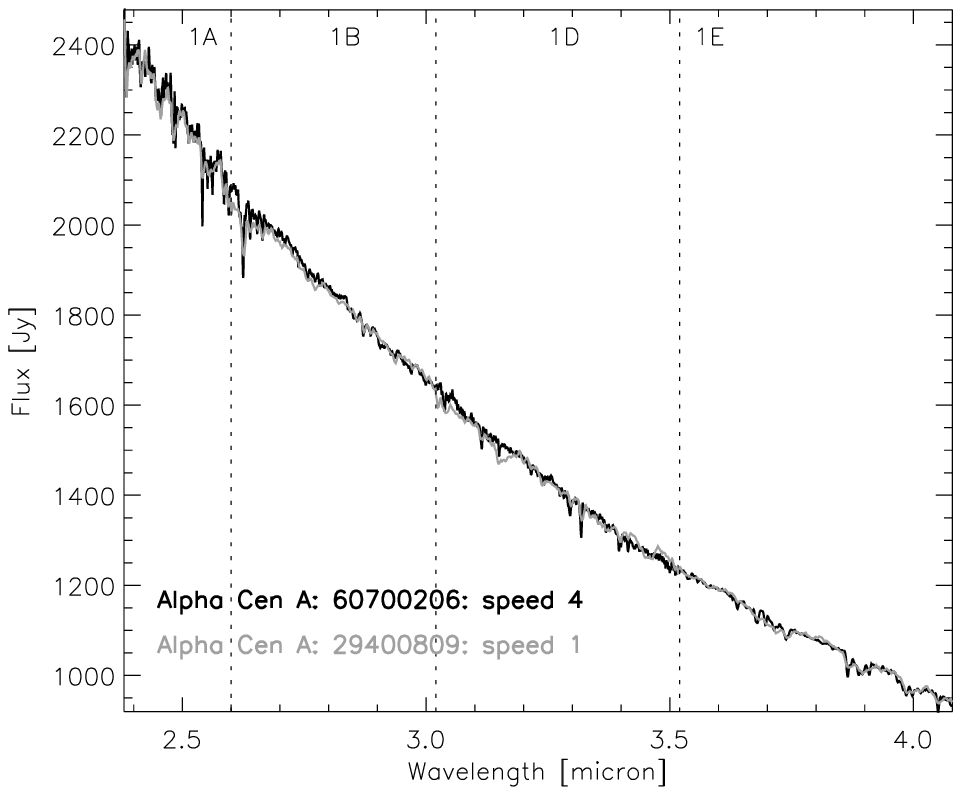}}}
\caption{\label{acenvers}Comparison between the AOT01 speed-4
observation of $\alpha$ Cen A (revolution 607) and the speed-1
observation (revolution 294). The data of the speed-1 observation
have been multiplied by a factor 1.16. A coloured version of this plot
is available in the Appendix as Fig.\ \ref{acenverscol}.}
\end{center}
\end{figure}

Alpha Cen A has also been observed during revolution 294 with the AOT01 speed-1
option. The pointing offsets were dy = $-0.797''$ and dz = $-0.832''$. Also for
this observation, the contribution of the K main-sequence companion of $\alpha$
Cen A is negligible.  The data of both band 1B and band 1E have been multiplied
by a factor 1.01. The relative features match quite well taking the
uncertainties of the speed-1 observation into account. There is, however, a
difference in absolute flux level of 16\,\% (Fig.\ \ref{acenvers}). In
revolution 294, the activation of the scientific measurements was
started later than usual because of the time allocated for the
Delta-V manoeuvre and the measurement of the superfluid He mass. These
two activities
may have influenced the quality of the speed-1 observation.
As will be argued in a subsequent article in this series --- where we
will confront the obtained synthetic spectra with the templates of
Cohen \citep{Cohen1992AJ....104.2030C,
Cohen1995AJ....110..275C, Cohen1996AJ....112.2274C,
Witteborn1999AJ....117.2552W} ---
 it is reasonable to assume that the absolute flux level
of the speed-4 observation is somewhat too high. Since the
absolute flux accuracy is quoted to be $\sim 10\,\%$, this 16\,\%
flux-difference is still within the quoted error bar.

\subsubsection{Comparison between the ISO-SWS spectrum and the synthetic
spectrum (Fig.\ \ref{acen})}

\begin{figure}[h!]
\begin{center}
\resizebox{0.5\textwidth}{!}{\rotatebox{90}{\includegraphics{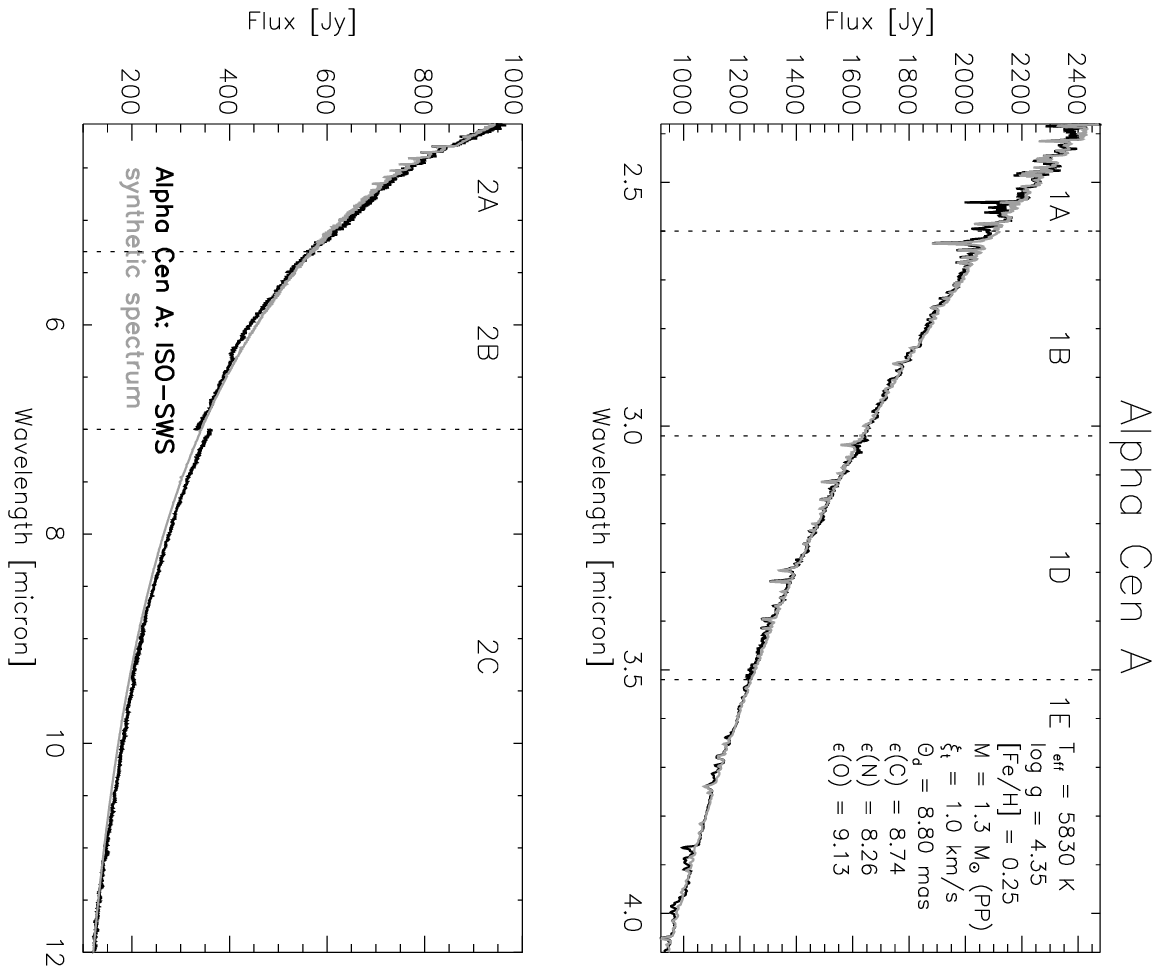}}}
\caption{\label{acen} Comparison between band 1 and band 2 of the
ISO-SWS data of $\alpha$ Cen A (black) and the synthetic spectrum
(grey) with stellar parameters \teff\ = 5830\,K, $\log$ g = 4.35, M
= 1.3\,\Msun, [Fe/H] = 0.25, \vt\ = 1.0\,\kms, $\varepsilon$(C)
= 8.74, $\varepsilon$(N) = 8.26, $\varepsilon$(O) = 9.13 and \ad\
= 8.80\,mas. A coloured version of this plot
is available in the Appendix as Fig.\ \ref{acencol}.}
\end{center}
\end{figure}

As discussed in \citetalias{Decin2000b}, it is quite difficult
to pin down the fundamental parameters of $\alpha$ Cen A from the SWS-spectrum,
due to the absence of molecular features. Therefore, the parameters
found by \citet{Neuforge1997A&A...328..261N} were used to calculate the
corresponding synthetic spectrum. Subsequently, the angular diameter, radius,
mass and luminosity were derived. This resulted in the following parameters:
\teff\ = $5830 \pm 30$\,K, $\log$ g = $4.35 \pm 0.05$, \vt\ = $1.0 \pm
0.1$\,\kms, [Fe/H] = $0.25 \pm 0.02$, $\varepsilon$(C) = $8.74 \pm 0.05$,
$\varepsilon$(N) = $8.26 \pm 0.09$, $\varepsilon$(O) = $9.13 \pm
0.06$, $\pi = 737.0 \pm 2.6$\,mas, \ad\ = $8.80 \pm
0.51$\,mas, R = $1.27 \pm 0.08$\,\Rsun, M$_g$ = $1.35 \pm
0.22$\,\Msun\ and L = $1.7 \pm 0.2$\,\Lsun, with deviation estimating parameters
from the Kolmogorov-Smirnov test (see Paper~I) being
$\beta_{\mathrm{1A}}$ = 0.043, $\beta_{\mathrm{1B}}$ = 0.036,
$\beta_{\mathrm{1D}}$ = 0.063, $\beta_{\mathrm{1E}}$ = 0.024.

Looking to the relative contribution of the different chemical species
\citep[see Fig.\ 3.1 in][]{Decinthesis} and to the Atmospheric Trace
Molecule Spectroscopy (ATMOS) spectrum of the Sun
\citep{Geller1992, Gunson1996GeoRL..23.2333G}, it is obvious that the atoms
dominate the infrared spectrum of $\alpha$ Cen A,
although the CO fundamental and first-overtone bands start arising around
4.4\,\mic\ and 2.4\,\mic\ respectively. As described in Paper II,
problems with inaccurate oscillator strengths of the atomic lines in
the infrared in the line list of \citet{Hirhor95}
caused quite some discrepancies between the
ISO-SWS and synthetic spectra. By using the identifications given
by \citet{Geller1992} for the ATMOS solar spectrum, the strongest
contributors to the most prominent features were identified, with the
strongest lines originating from Fe, Al, Si and Mg transitions (see,
e.g., Fig.\ 5 in Paper II).
The largest difference between the ISO-SWS spectrum of $\alpha$
Cen A and the rebinned ATMOS spectrum of the Sun occurs around
2.4\,\mic. As discussed in \citetalias{Decin2000b}, the origin of
this problem is situated in the problematic RSRF of band 1A in
this wavelength region. The problems with the computation of the
Humphreys lines near the Humphreys ionisation edge result in
$\beta_{\mathrm{1D}}$ being higher than the maximum acceptable
value for this sub-band as given in Table 3 in Paper I.

\subsubsection{Comparison with other published stellar parameters}

\begin{table*}[t!]
\caption{\label{litacena}Literature study of $\alpha$ Cen A, with the effective
temperature \teff\ given in K, the mass M in \Msun, the microturbulent velocity
$\xi_t$ in km/s, the angular diameter $\theta_d$ in mas, the luminosity L
in L$_{\odot}$ and the radius R in R$_{\odot}$. Angular diameters deduced from
direct measurements (e.g from interferometry) are written in italic, while
others (e.g.\ from spectrophotometric comparisons) are written
upright. Assumed or adopted values are given between parentheses.
The results of this research are mentioned in the last line. Only
the error bars on the {\it deduced} parameters are given. A short
description of the methods and/or data used by the
several authors can be retrieved from http://www.ster.kuleuven.ac.be/\~\,$\!$leen/artikels/ISO3/appendix.ps.}
\begin{center}
\setlength{\tabcolsep}{0.6mm}
\tiny
\begin{tabular}{ccccccccccccl} \hline
\rule[-3mm]{0mm}{8mm}  \teff & $\log$ g & M & $\xi_t$
&  [Fe/H] & $\varepsilon$(C) & $\varepsilon$(N) & $\varepsilon$(O)
& $\theta_d$ & L & R & Ref.\\ \hline \hline
\rule[-0mm]{0mm}{5mm} 5800 & & & & & &  & & $8.62 \pm 0.23$ & & & 1. \\
$5800 \pm 100$ & & $1.11 \pm 0.03$ & & & &  & & & $1.51 \pm 0.06$ & & 2. \\
 & & 1.10 & & & &  & & & & & 3. \\
$5750 \pm 30$ & $4.38 \pm 0.07$ & 1.08 & 1.0 & $0.28 \pm 0.15$ & &  & & & & & 4. \\
 & & $1.085 \pm 0.010$ & & & &  & & & & & 5. \\
$5820 \pm 50$ & $4.40 \pm 0.05$ & & $1.54 \pm 0.08$ & $0.20 \pm 0.04$ & &  & & & & & 6. \\
$5770 \pm 20$ & & & & & &  & & & & 1.23 & 7. \\
($5770$) & 4.5 & & (1.0) & $0.22 \pm 0.15$ & & & & & & & 8. \\
(5750) & $4.42 \pm 0.11$ & & (1.5) & (0.28) & &  & & & & & 9. \\
$5700 \pm 75$ & & & & & &  & & & 1.446 & $1.23 \pm 0.03$ & 10. \\
$5710 \pm 25$ & $4.27 \pm 0.20$ & (1.085) & $1.0 \pm 0.2$ & $0.12 \pm 0.06$ &
$8.77 \pm 0.04$ &  & $8.98 \pm 0.06$ & & & 1.26 & 11. \\
$5834 \pm 140$ & & & & & &  & & & & & 12. \\
$5800 \pm 20$ & $4.31 \pm 0.02$ & (1.085) & (1.0) & $0.22 \pm 0.02$ & &  &
& & & & 13. \\
5760 & & & & & &  & & 8.52 & & & 14. \\
5710 & & (1.085) & & & &  & & & $1.33 \pm 0.11$ & $1.17 \pm 0.05$ & 15. \\
& 4.3 & 1.4 & & & & & & & & & 16. \\
(5770) & (4.29) &  & {\tiny{$1.28 \pm 0.05 | 1.17 \pm 0.06$}} & 0.00 & & & &
& & &  17$^a$. \\
(5770) & (4.29) & & {\tiny{1.48-1.06 $|$ {\phantom{$1.17 \pm 0.06$}}}} & 0.30 & &  & & & &  & 17$^b$. \\
$5830 \pm 30$ & $4.34 \pm 0.05$ & ($1.085$) & $1.09 \pm 0.11$ & $0.25
\pm 0.02$ & $8.74 \pm 0.05$ &  $8.26 \pm 0.09$ & $9.13 \pm 0.06$ & & 1.5 & & 18. \\
 & & $1.160 \pm 0.031$ & & & &  & & & & & 19. \\
\rule[-3mm]{0mm}{3mm} ($5830$) & ($4.35$) & $1.35 \pm 0.22$ & ($1.0$)
& ($0.25$) &
($8.74$) &  ($8.26$) & ($9.13$) & $8.80 \pm 0.51$ & $1.7 \pm 0.2$ &
$1.27 \pm 0.08$ & 20. \\
\hline 
\end{tabular}
\end{center}
{\footnotesize{
1.\ \citet{Blackwell1977MNRAS.180..177B};
2.\ \citet{Flannery1978ApJ...221..175F};
3.\ \citet{Kamper1978AJ.....83.1653K};
4.\ \citet{England1980MNRAS.191...23E};
5.\ \citet{Demarque1986ApJ...300..773D};
6.\ \citet{Smith1986ApJ...311..843S};
7.\ \citet{Soderblom1986A&A...158..273S};
8.\ \citet{Abia1988A&A...206..100A};
9.\ \citet{Edvardsson1988A&A...190..148E};
10.\ \citet{Volk1989AJ.....98.1918V};
11.\ \citet{Furenlid1990ApJ...350..827F};
12.\ \citet{McWilliam1990ApJS...74.1075M};
13.\ \citet{Chmielewski1992A&A...263..219C};
14.\ \citet{Engelke1992AJ....104.1248E};
15.\ \citet{Pottasch1992A&A...264..138P};
16.\ \citet{Popper1993ApJ...404L..67P};
17.\ \citet{Gadun1994AN....315..413G};
18.\ \citet{Neuforge1997A&A...328..261N};
19.\ \citet{Pourbaix1999A&A...344..172P};
20.\ present results}}
\end{table*}

\begin{itemize}
\item{\underline{Assumed parameters:}\\
It has to be noted that the fundamental stellar parameters (\teff,
$\log$ g, \vt, [Fe/H], $\varepsilon$(C), $\varepsilon$(N),
$\varepsilon$(O)), determined by several authors using different
methods, are in excellent agreement (see Table \ref{litacena}). Since the
most up-to-date spectroscopic analysis based on high-resolution and
high signal-to-noise spectra of $\alpha$ Cen A was performed by
\citet{Neuforge1997A&A...328..261N}, their derived parameters were
used as input parameters.}
\item{\underline{Deduced parameters:}\\
The angular diameter
deduced from the AOT01 speed-4 observation of $\alpha$ Cen A
(revolution 607), is somewhat larger than the values obtained by
other indirect methods, but is still within the error bars of the
other values. The origin of the difference may be too high a flux level of the
ISO-SWS AOT01 speed-4 observation (see Sect.\ \ref{compobsacena}). As a
consequence, the stellar radius, gravity-inferred mass and luminosity are also
somewhat larger than the other values listed in Table
\ref{litacena}. Different methods were used by different authors to
estimate the radius of $\alpha$ Cen A: using \teff\ and L
\citep[e.g.,][]{Soderblom1986A&A...158..273S,
Furenlid1990ApJ...350..827F}, using \ad\ and $\pi$
\citep[e.g.,][]{Volk1989AJ.....98.1918V} or using the p-mode
oscillations found in $\alpha$ Cen A
\citep[e.g.,][]{Pottasch1992A&A...264..138P}. The same can be said for
the luminosity, where \citet{Flannery1978ApJ...221..175F} have used
different broad-band systems and narrow-band photometric indices to
estimate the luminosity; \citet{Volk1989AJ.....98.1918V} and
\citet{Pottasch1992A&A...264..138P} have used the assumed \teff\ and
deduced
R-values. The most quoted mass value for $\alpha$ Cen A is the one
deduced by \citet{Demarque1986ApJ...300..773D} (M $= 1.085 \pm
0.010$\,\Msun). \citet{Pourbaix1999A&A...344..172P} found for the
first time an agreement between astrometric and spectroscopic mass
ratio. Their value for the mass (M $= 1.160 \pm 0.031$\,\Msun) is in
better agreement with a mass estimated from the evolutionary tracks of
\citet{Girardi2000A&AS..141..371G} (for \teff\ = 5830\,K and L =
1.7\,\Lsun, we found M $= 1.02 \pm 0.20$\,\Msun) than our
gravity-inferred mass of $1.30 \pm 0.21$\,\Msun.}
\end{itemize}


\subsection{$\alpha$ Car \rm{: AOT01, speed 4, revolution 729}}\label{disacar}

\subsubsection{Some specific calibration details}

Only the band-1A flux had to be multiplied with a factor larger
than 1.02, although this is still smaller than the mean
band-border ratio between band 1A and band 1B for that revolution (see
Fig.\ 1 in Feuchtgruber, 1998).

\subsubsection{Comparison between the ISO-SWS spectrum and the synthetic
spectrum (Fig.\ \ref{acar})}

\begin{figure}[h!]
\begin{center}
\resizebox{0.5\textwidth}{!}{\rotatebox{90}{\includegraphics{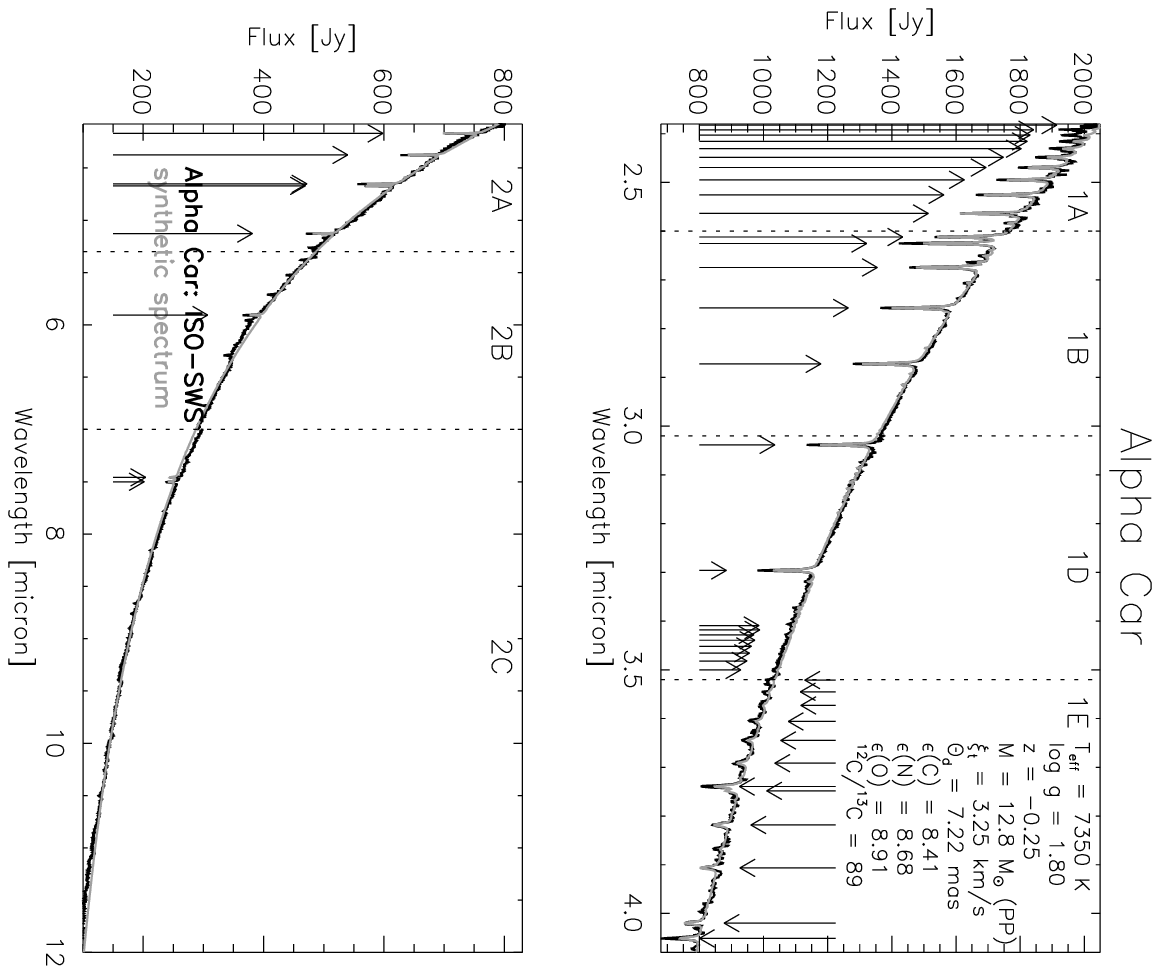}}}
\caption{\label{acar} Comparison between band 1 and band 2 of the
ISO-SWS data of $\alpha$ Car (black) and the synthetic spectrum
(grey) with stellar parameters \teff\ = 7350\,K, $\log$ g = 1.80, M
= 12.8\,\Msun, [Fe/H] = $-0.25$, \vt\ = 3.25\,\kms,
$\varepsilon$(C) = 8.41, $\varepsilon$(N) = 8.68, $\varepsilon$(O)
= 8.91 and \ad\ = 7.22\,mas. Hydrogen lines are indicated by
arrows.
A coloured version of this plot is available in the Appendix as Fig.\
\ref{acarcol}.}
\end{center}
\end{figure}

Using the stellar parameters \teff\ = $7350$\,K, $\log$ g = $1.80$,  \vt\ = $3.25$\,\kms, [Fe/H] = $-0.24$, $\varepsilon$(C) = $8.41$,
$\varepsilon$(N) = $8.68$, $\varepsilon$(O) = $8.91$
\citep{Desikachary1982MNRAS.201..707D} and $\pi = 10.43 \pm
0.53$\,mas results in \ad\ = $7.22 \pm
0.42$\,mas, R = $74.39 \pm 5.76$\,\Rsun, M$_g$ =
$12.80^{+24.95}_{-6.35}$\,\Msun\ and L = $14573 \pm 2268$\,\Lsun\ with
deviation estimating parameters
$\beta_{\mathrm{1A}}$ = 0.091, $\beta_{\mathrm{1B}}$ = 0.045,
$\beta_{\mathrm{1D}}$ = 0.091, $\beta_{\mathrm{1E}}$ = 0.041.

Just as for $\alpha$ Cen A, the large $\beta$-values from the
Kol\-mo\-go\-rov-Smirnov test may be explained by the problematic
prediction of the atomic lines and especially
the hydrogen lines (which dominate the spectral signature) and the
noise. The more pronounced discrepancy
visible at the beginning of band 1A is due to calibration problems (RSRF). The
lower gravity of $\alpha$ Car with respect to the other {\it{warm}} stars is
reflected in the smaller broadening of the  hydrogen lines. Despite
this lower gravity, all but one sub-band are rejected by the
Kolmogorov-Smirnov test.

\subsubsection{Comparison with other published stellar parameters}

\begin{table*}[t!]
\caption{\label{litacar}See caption of Table \ref{litacena}, but now
for $\alpha$ Car.}
\begin{center}
\setlength{\tabcolsep}{.6mm}
\tiny
\begin{tabular}{cccccccccccl} \hline 
\rule[-3mm]{0mm}{8mm}  \teff & $\log$ g & M & $\xi_t$
&  [Fe/H] & $\varepsilon$(C) & $\varepsilon$(N) & $\varepsilon$(O)
& $\theta_d$ & L & R & Ref.\\ \hline \hline
\rule[-0mm]{0mm}{5mm}  & & & & & &  & & ${\mathit{6.6 \pm 0.8}}$ & & & 1. \\
$7460 \pm 460$ & & & & & &  & & (${\mathit{6.6}}$) & & & 2. \\
$7206 \pm 173$ & & & & & &  & & $7.08 \pm 0.19$ & & & 3. \\
7420 & & & & & &  & & & & & 4. \\
$7500 \pm 200$ & $1.85 \pm 0.30$ & & $2.7 \pm 1.0$ & $0.35 \pm 0.15$ & &  & & & & & 5. \\
$7346 \pm 150$  & & & & & &  & & $6.81 \pm 0.20$ & & & 6. \\
$7350 \pm 300$ & $1.80 \pm 0.50$ & & $3.25 \pm 0.25$ & $-0.24 \pm 0.04$ & $8.41
\pm 0.10$ & $8.68 \pm 0.05$ & $8.91 \pm 0.10$ & & & & 7. \\
$7400 \pm 150$ & $1.9 \pm 0.2$ & & $4.5|6.0$ & & &  & & & & & 8. \\
($7400$) & ($1.9$) & $8 \pm 1.5$ & (4.5) & $-0.10$ & 8.32 &  & & & 7500 & 53 & 9. \\
7320 -- 7900 & $1.8 \pm 0.2$ & & 3.0 & & &  & & & & & 10$^a$. \\
$7500 \pm 200$ & $1.5 \pm 0.3$ & & $3.5 \pm 0.5$ & $-0.07$ & 8.27 & 8.24 & 8.69 & & & & 10$^b$. \\
 & & & & & &  & & ${\mathit{6.6 \pm 0.8}}$ & & & 11. \\
$7260 \pm 200$ & $1.83 \pm 0.30$ & & & &  & & & & & &
12$^a$. \\
($7400$) & ($1.9$) & 7.94 & $4.5|5.7$ & $-0.12$ & 7.33 &  & & & & 52.48 & 12$^b$. \\
(7500) & 1.2 & & 3.0 & $+0.08 \pm 0.10$ & &  & & & & & 13. \\
$7460 \pm 460$ & & & & & &  & & & ($1795$) & $ 25.5 \pm 3.9$ & 14. \\
$7298 \pm 150$ & & & & & &  & & & & & 15. \\
$7500 \pm 100$ & $1.2 \pm 0.2$ & 8-9 & $2.8 \pm 0.2$ & 0.00 & &  & & & & &
16. \\
(7500) & (1.5) & 12.6 & & & &  & & & & & 17. \\
7350 & 1.80 & 8 & 1.99-4.27 $|$ 2.40-5.44 & $-0.3 \pm
0.11$ & &  & & & & 53 & 18. \\
$7500 \pm 200$ & $1.5 \pm 0.3$ & & $2.5 \pm 0.5$ & $0.06 \pm 0.15$ & $8.41 \pm
0.10$ &   & $8.63 \pm 0.2$ & & & & 19. \\
$7520 \pm 460$ & $<1.5$ & & & & &  & & (${\mathit{6.6}}$) & & & 20. \\
\rule[-3mm]{0mm}{3mm}($7350$) & ($1.80$) & $12.8^{+24.95}_{-6.35}$ &
($3.25$) & ($-0.25$) &  ($8.41$) & ($8.68$) & ($8.91$) &  $7.22 \pm 0.42$ &
 $14573 \pm 2268$ & $74.39 \pm 5.76$ & 21. \\
\hline 
\end{tabular}
\end{center}
{\footnotesize{
1.\ \citet{Brown1974MNRAS.167..121B};
2.\ \citet{Code1976ApJ...203..417C};
3.\ \citet{Blackwell1977MNRAS.180..177B};
4.\ \citet{Linsky1978ApJ...220..619L};
5.\ \citet{Luck1979ApJ...232..797L};
6.\ \citet{Blackwell1980A&A....82..249B};
7.\ \citet{Desikachary1982MNRAS.201..707D};
8.\ \citet{Boyarchuk1983Ap.....18..228B};
9.\ \citet{Lyubimkov1984Ap.....19..385L};
10.\ \citet{Luck1985ApJ...298..782L};
11.\ \citet{DiBenedetto1987A&A...188..114D};
12.\ \citet{Russell1989ApJS...70..865R};
13.\ \citet{Spite1989A&A...210...25S};
14.\ \citet{Volk1989AJ.....98.1918V};
15.\ \citet{McWilliam1990ApJS...74.1075M};
16.\ \citet{Achmad1991A&A...249..192A};
17.\ \citet{ElEid1994A&A...285..915E};
18.\ \citet{Gadun1994AN....315..413G};
19.\ \citet{Hill1995A&A...293..347H};
20.\ \citet{Smalley1995A&A...293..446S};
21.\ present results}}
\end{table*}

\begin{itemize}
\item{\underline{Assumed parameters:}\\
From Table \ref{litacar}, it is clearly apparent that a gravity
determined in a spectroscopic way (i.e.\ by requiring that the
abundance of neutral and ionised lines yield the same abundance)
is usually lower than a photometric gravity (i.e.\ determined from
photometric colours, or by using the well-known relation between
g, \teff, M and L, where the mass has been determined by locating
the stellar object on theoretical evolutionary tracks). It is well
known that the accuracy of spectroscopic determinations of
gravities from ionisation equilibria or molecular equilibria for
individual stars is not very good for giants \citep[cf.,
e.g.,][]{Trimble1981QJRAS..22..361T, Brown1983ApJ...265L..93B,
Smith1985ApJ...294..326S}. Therefore, the values given by
\citet{Desikachary1982MNRAS.201..707D} were adopted as atmospheric
parameters for $\alpha$ Car. }
\item{\underline{Deduced parameters:}\\
The angular diameter derived from the ISO-SWS data is larger than
the two values based on the InfraRed Flux Method
\citep[IRFM,][]{Blackwell1977MNRAS.180..177B,
Blackwell1980A&A....82..249B} in Table \ref{litacar}.
\citet{Napiwotzki1993A&A...268..653N} quoted that temperatures
deduced by using the IRFM are {\it too low} by 1.6 -- 2.8\,\%,
corresponding to angular diameters which are too large by 3.5 --
5.9\,\%. When inspecting the other stars of the sample which have
been analysed by means of the IRFM, this trend however is not
visible. The reason for the discrepancy in angular diameter may be
the problematic determination of the continuum in the spectrum of
a {\it{warm}} star \citepalias{Decin2000b}. This larger angular
diameter however can not explain the difference in radius and
luminosity seen between our results and other literature values.
\citet{Lyubimkov1984Ap.....19..385L} have estimated a mass value
of $8 \pm 1.5$\,\Msun\ for $\alpha$ Car from the evolutionary
tracks of \citet{Becker1981ApJS...45..475B} in the $\log$ \teff -
$\log$ g plane. Also \citet{Russell1989ApJS...70..865R} have
determined the mass from evolutionary tracks. Together with the
gravity mentioned by \citet{Boyarchuk1983Ap.....18..228B}, this
results in the stellar radius of $\sim 53$\,\Rsun. Using $V-I
= 0.44$ \citep{Johnson1966CoLPL...4...99J}, the bolometric
correction BC$_I$ from \citet{Bessell1998A&A...333..231B} and the
Hipparcos' parallax, we obtain  $\rm{M_{bol}} = -5.68 \pm 0.12$ (L
= $14723 \pm 1654$\,\Lsun), in agreement with our \teff\ based
luminosity. From the evolutionary tracks of
\citet{Claret1995A&AS..114..549C} we estimate a somewhat higher
mass value being $10.5 \pm 0.5$\,\Msun. When our gravity-inferred
mass of 12.8\,\Msun\ would have been used by
\citet{Lyubimkov1984Ap.....19..385L} and
\citet{Russell1989ApJS...70..865R}, a stellar radius of 67\,\Rsun\
and a luminosity of 12000\,\Lsun\ would have been deduced, in
close agreement with our results. \citet{Volk1989AJ.....98.1918V}
found a luminosity of $1795 \pm 71$\,\Lsun. This value was
determined from the effective temperature, which was deduced
directly from the literature values of angular diameter
measurements, total-flux observations (from the literature) and
the parallax from \citet{Hoffleit1982}. The parallax value
mentioned by \citet{Hoffleit1982} ($\pi = 28$\,mas) is however a
factor 2.68 higher than the value given by the Hipparcos
catalogue. Using the Hipparcos' parallax would increase the
stellar radius by the same amount --- resulting in R = 68\,\Rsun\
--- and the luminosity by a factor 7.2 --- giving L $\approx
13000$\,\Lsun, both values now being in good agreement with our
deduced values.}
\end{itemize}


\subsection{$\beta$ Leo \rm{: AOT01, speed 3, revolution 189}}\label{disbleo}

\subsubsection{Some specific calibration details}

$\beta$ Leo is one of the few stars of our sample for which no
speed-4 observation has been obtained. The signal-to-noise is
therefore smaller than for the other {\it{warm}} stars. The shape
of the spectrum in band 2 (after application of a standard
calibration procedure) is very suspicious.  A quick-look
at the SPD (\,=\,Standard Processed Data, which gives the signal as a
function of time) reveals immediately the origin of the
problem. A photometric check is taken between the measurements
with aperture 1 and aperture 2 and between the measurements with
aperture 2 and aperture 3, i.e.\ just before the up scan of band
2A, and before the up scan of band 2C (Fig.\ \ref{spdbleo}). The
calibration source is, however, much brighter than $\beta$ Leo,
resulting in a strong memory effect.  Only the dark current before
each measurement (at a certain aperture) has the same reset time
as the measurement itself and could be used to subtract the dark
current. Since the dark current before the measurement with
aperture 3 (band 2C) is strongly affected by memory effects
arising from the photometric check, the mean flux value of the
dark current is higher than the flux values of the down scan
beyond 9\,\mic. Therefore only the up scan of band 2C is used, and
the fact is taken into account that memory effects are destroying
the reliability of this up scan. In order to correct for this too
high dark current, the data of band 2C are shifted upwards by 3.5
Jy (Table 3 in Paper~II). The other factors used to combine the different
sub-bands may also be found in Table 3 in Paper~II.

\begin{figure}[h!]
\begin{center}
\resizebox{0.5\textwidth}{!}{\rotatebox{90}{\includegraphics{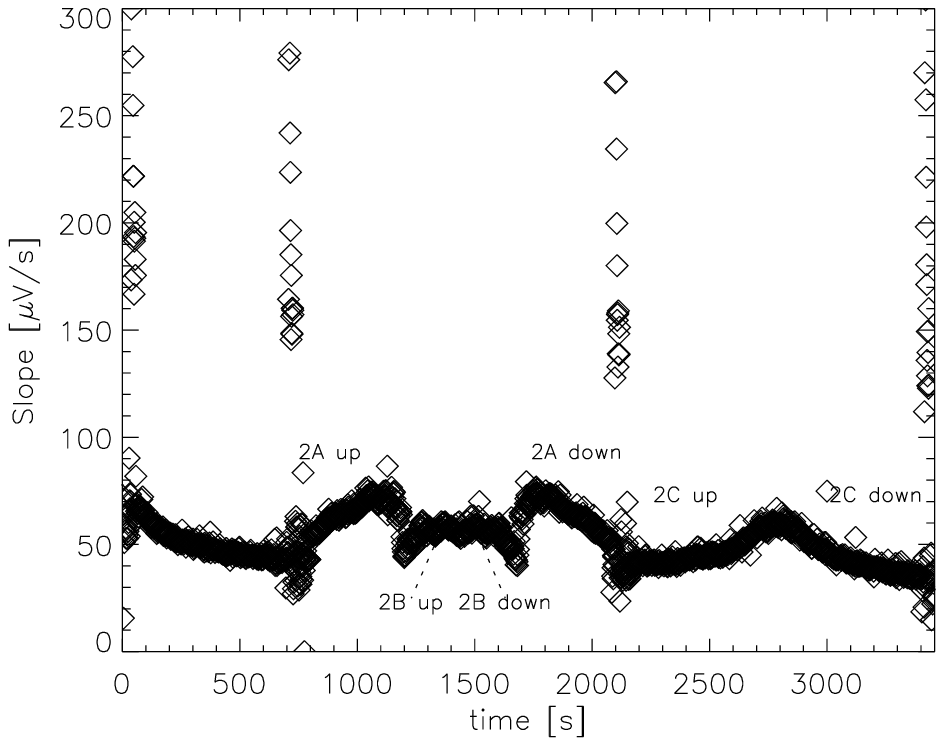}}}
\caption{\label{spdbleo}The slope of detector 13 of band 2 of the
AOT01 observation of $\beta$ Leo (revolution 189) is plotted against
time. The up and down scans of the different sub-bands are
indicated. Photometric checks are taken between aperture 1 and aperture 2 and
between aperture 2 and aperture 3.}
\end{center}
\end{figure}

\subsubsection{Comparison with other AOT01 observations}

$\beta$ Leo has also been observed in revolution 040 during PV
(Performance Verification) with the AOT01 speed-1 option. The
pointing offsets were dy = $-0.737''$ and dz = $0.913''$. Due to
the triangular shape of the instrumental beam profile in the
cross-dispersion direction of SWS, a pointing offset in the cross-dispersion direction causes a higher
signal loss than in the dispersion-direction.
For this observation, a photometric check was only taken before
the up scan of band 2A and after the down scan of band 2C. So,
only the up scan of band 2A can be affected by memory effects
originating from the photometric check. The down scan of band 2A
(and band 1D) displays however a signal jump 640 s after the start
of the observation (Fig.\ \ref{spd2bleo}). The origin of such
signal jumps is at the moment unclear \citep[][p.~66]{Leech2002}.
The jumps in bands 1 and 2 are similar to each other, but
different from jumps in band 3. They can be negative or positive,
and there seems to be a relation between the signal jump and a
residual pulse effect after reset. It is recommended to adjust the
baseline of the affected portion to the pre-jump baseline.
For reason of safety, these data have been flagged as `no-data'.
In order to obtain a smooth spectrum, the data of the sub-bands
1A, 1B and 1E of this speed-1 observation have been multiplied by
a factor 1.01.

\begin{figure}[h!]
\begin{center}
\resizebox{0.5\textwidth}{!}{\rotatebox{90}{\includegraphics{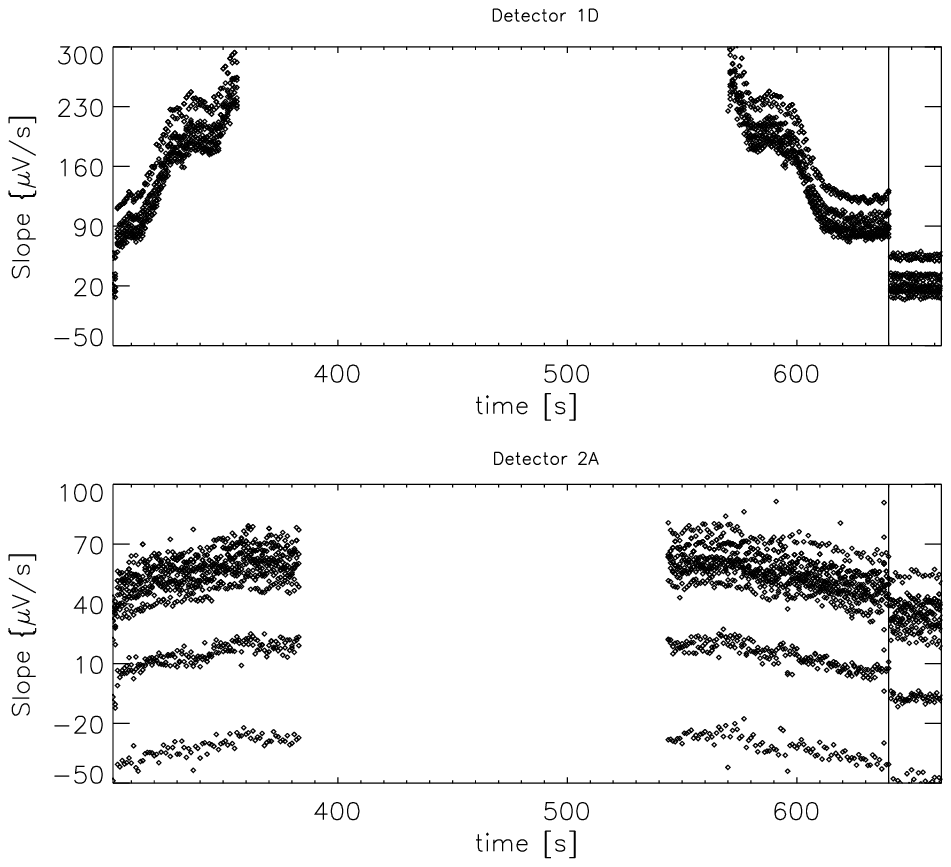}}}
\caption{\label{spd2bleo}The slope of the detectors in band 1D and
band 2A of the AOT01 speed-1 observation of $\beta$ Leo (revolution 040)
are plotted against time. The signal jump is indicated by the
vertical line.}
\end{center}
\end{figure}

The photometric flux level of this observation (revolution 040) is
about 5\,\% lower than the AOT01 speed-4 observation taken during
revolution 189. A few differences between the two observed spectra are somewhat
more pronounced, e.g. around 2.42\,\mic, 2.58\,\mic, 3.8\,\mic\ (see
Fig.\ \ref{bleovers}). Inaccuracies in the speed-1 observations --- clearly
visible from the comparison between up and down scan --- originate
from these discrepancies.

\begin{figure}[h!]
\begin{center}
\resizebox{0.5\textwidth}{!}{\rotatebox{90}{\includegraphics{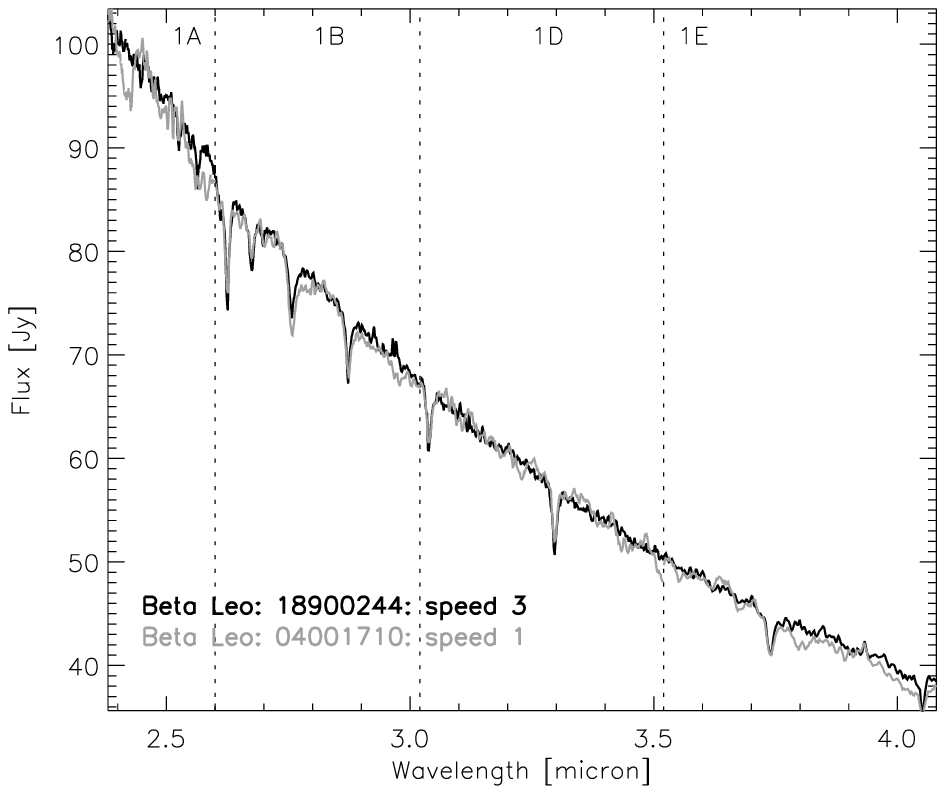}}}
\caption{\label{bleovers}Comparison between the AOT01 speed-3
observation of $\beta$ Leo (revolution 189) and the speed-1
observation (revolution 040). The data of the speed-1 observation
have been multiplied by a factor 1.05. A coloured version of this plot
is available in the Appendix as Fig.\ \ref{bleoverscol}.}
\end{center}
\end{figure}

\subsubsection{Comparison between the ISO-SWS spectrum and the synthetic
spectrum (Fig.\ \ref{bleo})}

\begin{figure}[h!]
\begin{center}
\resizebox{0.5\textwidth}{!}{\rotatebox{90}{\includegraphics{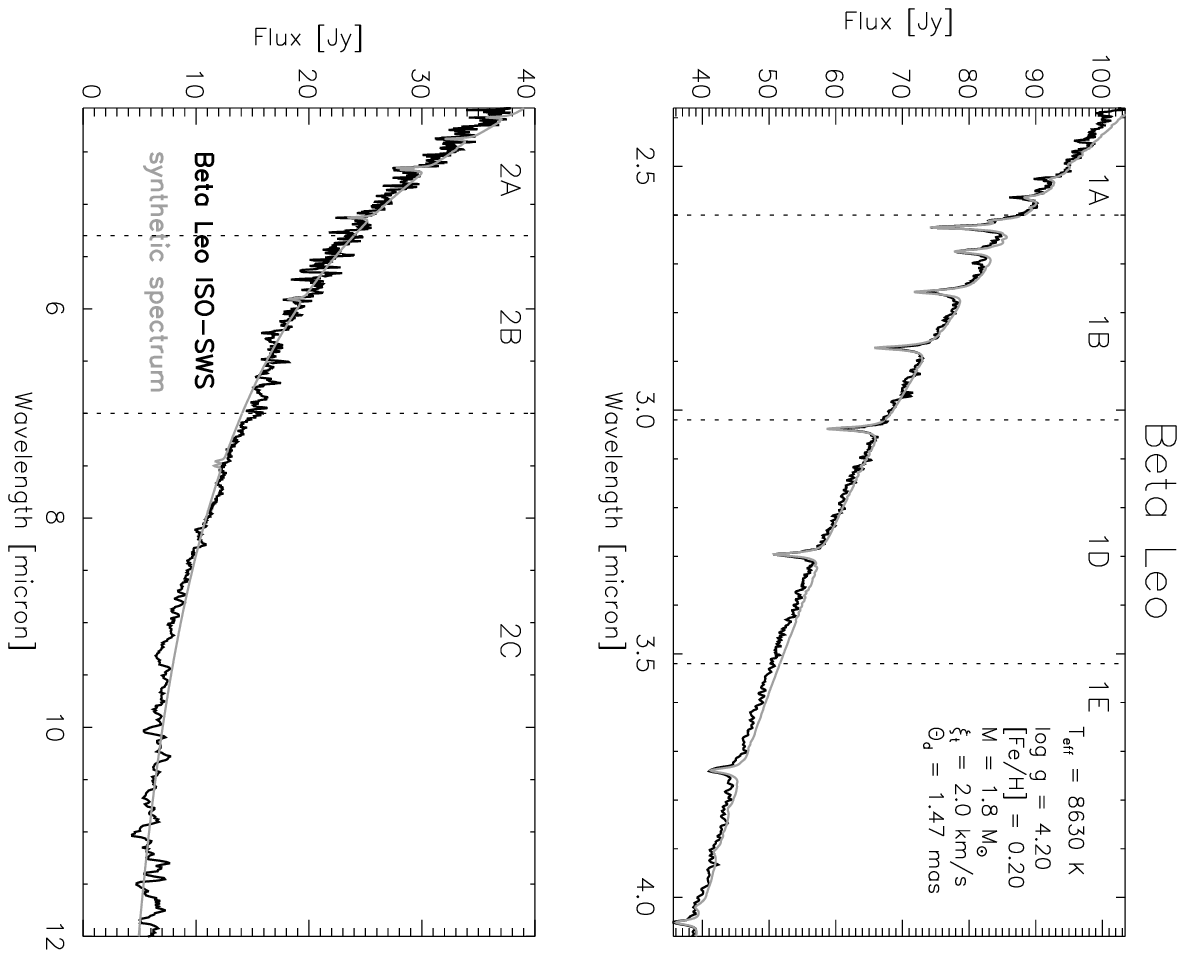}}}
\caption{\label{bleo} Comparison between band 1 and band 2 of the
ISO-SWS data of $\beta$ Leo (black) and the synthetic spectrum
(grey) with stellar parameters \teff\ = 8630\,K, $\log$ g = 4.20, M
= 1.8\,\Msun, [Fe/H] = 0.20, \vt\ = 2.0\,\kms\ and \ad\ = 1.47\,mas. A
coloured version of this plot is available in the Appendix as Fig.\
\ref{bleocol}.}
\end{center}
\end{figure}

Good-quality published stellar parameters for $\beta$ Leo were found in
\citet{Holweger1995A&A...303..819H}. These authors
list as parameters: \teff\ = $8630$\,K, $\log$ g = $4.20$. Using
Str{\o}mgren photometry, \citet{Gardiner1999A&A...347..876G} obtained
${\rm{Fe/H]}} \approx +0.2$\,dex. A microturbulent velocity of
2\,\kms\ was assumed.
With an angular diameter deduced from the ISO-SWS spectrum \ad\ =
$1.47 \pm 0.09$\,mas, one obtains R = $1.75 \pm 0.11$\,\Rsun, M$_g$ =
$1.78 \pm 1.04$\,\Msun\ and L =
$15 \pm 3$\,\Lsun. The corresponding deviation estimating parameters are
$\beta_{\mathrm{1A}}$ = 0.060, $\beta_{\mathrm{1B}}$ = 0.028,
$\beta_{\mathrm{1D}}$ = 0.087, $\beta_{\mathrm{1E}}$ = 0.038.

The large $\beta$-values for $\beta$ Leo are not very surprising.
One first of all has to take into account the problems with the
hydrogen lines, which dominate the spectrum. Secondly, for band
1A, there is also a large discrepancy at the wavelengths were the
H5-23 and H5-22 lines emerge at the beginning of this band. A
problem with the RSRF is at the origin of this discrepancy
\citepalias{Decin2000b}. The large $\beta_{\mathrm{1D}}$-value is
arising from the problems nearby the Humphreys ionisation edge
\citepalias{Decin2000b}. The lower signal/noise ratio in a AOT01
speed-3 observation compared to the other observations also
contributes to larger $\beta$-values.

\subsubsection{Comparison with other published stellar parameters}\label{literbleo}

\begin{table*}[t!]
\caption{\label{litbleo}See caption of Table \ref{litacena}, but now
for $\beta$ Leo.}
\begin{center}
\tiny
\begin{tabular}{cccccccccccl} \hline
\rule[-3mm]{0mm}{8mm}  \teff & $\log$ g & M & $\xi_t$
&  [Fe/H] & $\varepsilon$(C) & $\varepsilon$(N) & $\varepsilon$(O)
& $\theta_d$ & L & R & Ref.\\ \hline \hline
\rule[-0mm]{0mm}{5mm} & & & & & &  & & ${\mathit{1.33 \pm 0.10}}$ & & & 1. \\
$8850 \pm 340$ & & & & & &  & & (${\mathit{1.33}}$) & & & 2. \\
 & & & & & &  & & ${\mathit{1.32}}$ & & 1.73 & 3. \\
8660 & & & & & &  & & & & 1.73 & 4$^a$. \\
8660 & & & & & &  & & & & 1.68 & 4$^b$. \\
8600 &  4.2 & & & & & & & & & & 5. \\
$9590 \pm 460$ & $4.27 \pm 0.15$ & 2.3 & & & &  & & $1.39 \pm 0.03$ & $25.1 \pm 7.2$ & $1.82 \pm 0.18$ & 6. \\
8850 & & & & & &  & & & & & 7. \\
8500 & 4.20 & & & 0.00 & &  & & & & & 8$^a$. \\
8640 & 4.37 & & & 0.50 & &  & & & & & 8$^b$. \\
8310 & 4.20 & & & 0.00 & &  & & & & & 8$^c$. \\
8260 & 4.37 & & & 0.50 & &  & & & & & 8$^d$. \\
8630 & 4.21 & & & & &  & & & & & 9. \\
$8870 \pm 350$ & $4.00 \pm 0.25$ & & & & &  & & (${\mathit{1.40}}$) & & & 10$^a$. \\
$8870 \pm 350$ & $4.10 \pm 0.30$ & & & & &  & & (${\mathit{1.33}}$) & & & 10$^b$. \\
$8720 \pm 300$ & & & & & &  & & & & & 11. \\
$8857 \pm 185$ & 4.0 & & & & &  & & $1.374 \pm 0.033$ & & & 12. \\
(8870) & (4.10) & & & 0.20 & & & & & & & 13. \\
\rule[-3mm]{0mm}{3mm}($8630$) & ($4.20$) & $1.78 \pm 0.46$ & (2.0) &
(0.00) & ($8.76$) & ($8.25$) & ($9.13$) & $1.47 \pm
0.09$ & $15 \pm 2$ & $1.75 \pm 0.11$ & 14. \\
\hline
\end{tabular}
\end{center}
{\footnotesize{
1.\ \citet{Brown1974MNRAS.167..121B};
2.\ \citet{Code1976ApJ...203..417C};
3.\ \citet{Moon1985Ap&SS.117..261M};
4.\ \citet{Moon1985MNRAS.217..305M};
5.\ \citet{Lester1986ApJS...61..509L};
6.\ \citet{Malagnini1990A&AS...85.1015M};
7.\ \citet{Napiwotzki1993A&A...268..653N};
8.\ \citet{Smalley1993A&A...271..515S};
9.\ \citet{Holweger1995A&A...303..819H};
10.\ \citet{Smalley1995A&A...293..446S};
11.\ \citet{Sokolov1995A&AS..110..553S};
12.\ \citet{Malagnini1997A&A...326..736M};
13.\ \citet{Gardiner1999A&A...347..876G};
14.\ present results}}
\end{table*}

\begin{itemize}
\item{\underline{Assumed parameters:}\\
From the detailed investigation by \citet{Smalley1993A&A...271..515S},
one may conclude that atmospheric parameters derived from
photometry, spectrophotometry and from H$\beta$ and H$\gamma$
profiles all agree quite well, provided that adequate opacities
and metallicities are used. Especially the values of the effective
temperature and gravity should be very reliable when determined
from photometry, and they should not be significantly affected by
uncertainties in the metallicity
\citep{Smalley1993A&A...271..515S}. From the research of
\citet{Gardiner1999A&A...347..876G}, one may however conclude that the
determination of the effective temperature and the gravity for this
star from Balmer line profiles \citep[as done e.g.\
by][]{Smalley1995A&A...293..446S} seems to be rather uncertain,
because $\beta$ Leo is located close to the maximum of the Balmer line width.
Therefore, the values found by
\citet{Holweger1995A&A...303..819H} --- who have used Str\"{o}mgren photometry
--- were taken as stellar parameters for the theoretical model and
corresponding synthetic spectrum. Since
\citet{Gardiner1999A&A...347..876G} estimated $\beta$ Leo to be
slightly overabundant (${\rm{[Fe/H]}} \approx +0.2$\,dex) from
Str\"{o}mgren photometry, this value was assumed for the
metallicity. For the microturbulent velocity a value of 2\,\kms\ was
assumed.}
\item{\underline{Deduced parameters:}\\
\citet{Malagnini1990A&AS...85.1015M, Malagnini1997A&A...326..736M}
have used a flux-fitting method which determines simultaneously the
effective temperature and the angular diameter. Their obtained values
agree well with the angular diameter deduced from the ISO-SWS spectrum.
The small differences in the luminosity and radius values mentioned by
\citet{Malagnini1990A&AS...85.1015M}
are partly due to the use of the parallax value of \citet{Hoffleit1982},
which is a factor 1.1 lower than the Hipparcos' parallax.
When we then compare our gravity-inferred mass with the mass estimated
from evolutionary tracks by \citet{Malagnini1990A&AS...85.1015M}, we
see that our deduced value is somewhat lower than their quoted
value. Using $\log$ \teff\ $= 3.936$ and $\log$ L $= 1.176 \pm 0.058$,
we estimated from the evolutionary tracks of
\citet{Girardi2000A&AS..141..371G} a mass of 1.9\,\Msun, which is thus somewhat
closer to the value of 2.3\,\Msun, given by
\citet{Malagnini1990A&AS...85.1015M}.}
\end{itemize}


\subsection{Sirius\rm{: AOT01, speed 4, revolution 689}} \label{dissirius}

\subsubsection{Some specific calibration details}

The speed-4 observation of Sirius only suffered from very small
pointing errors. Bands 1A and 1B were shifted upwards by 1\,\%, band
1E was shifted downwards by 0.5\,\%.

\subsubsection{Comparison with other AOT01 observations}

At the end of the ISO mission --- during revolution 868 --- Sirius was
once more observed using the AOT01 mode, but now with a higher
speed, resulting in a lower resolution and a lower signal-to-noise
ratio. The pointing offsets were negligible. Only the data of band 1A
were divided by a factor 1.01 to optimize the match
between the different sub-bands. The absolute-flux levels differ
however by 12\,\% (Fig.\ \ref{siriusvers}). With a quoted
absolute-flux accuracy of 10\,\% and the template of Cohen
\citep{Cohen1992AJ....104.1650C} being in absolute-flux level in
between these two observations this difference is not worrying.

\begin{figure}[h!]
\begin{center}
\resizebox{0.5\textwidth}{!}{\rotatebox{90}{\includegraphics{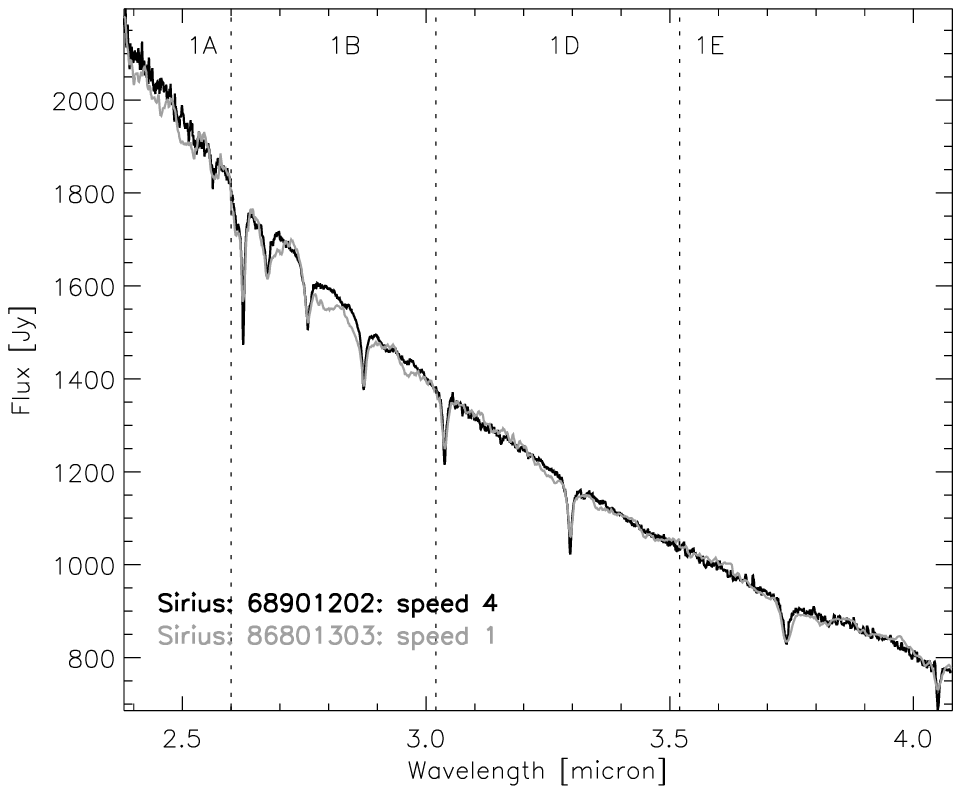}}}
\caption{\label{siriusvers}Comparison between the AOT01 speed-4
observation of $\alpha$ CMa (revolution 689) and the speed-1
observation (revolution 868). The data of the speed-1 observation
have been multiplied by a factor 1.12. A coloured version of this plot
is available in the Appendix as Fig.\ \ref{siriusverscol}.}
\end{center}
\end{figure}

Taking the lower signal-to-noise ratio of the speed-1 observation into
account, the features of the two observational spectra of Sirius do agree well.

\subsubsection{Comparison between the ISO-SWS spectrum and the synthetic
spectrum (Fig.\ \ref{sirius})}

\begin{figure}[h!]
\begin{center}
\resizebox{0.5\textwidth}{!}{\rotatebox{90}{\includegraphics{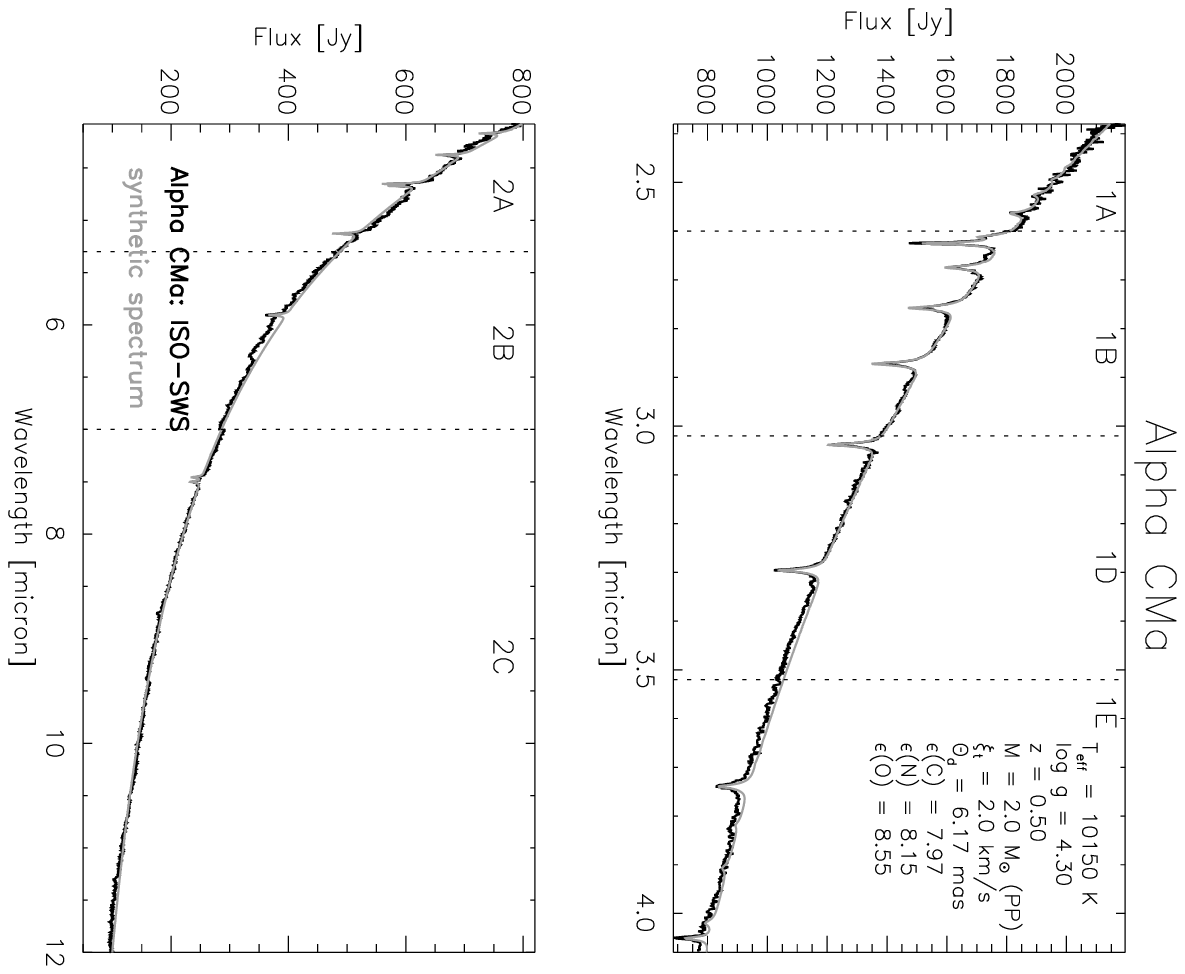}}}
\caption{\label{sirius} Comparison between band 1 and band 2 of
the ISO-SWS data of $\alpha$ CMa (black) and the synthetic
spectrum (grey) with stellar parameters \teff\ = 10150\,K, $\log$ g
= 4.30, M = 2.2\,\Msun, [Fe/H] = 0.50, \vt\ = 2.0\,\kms,
$\varepsilon$(C) = 7.97, $\varepsilon$(N) = 8.15, $\varepsilon$(O)
= 8.55 and \ad\ = 6.17\,mas. A coloured version of this plot is
available in the Appendix as Fig.\ \ref{siriuscol}.}
\end{center}
\end{figure}

For Sirius the effective temperature, the gravity and the
metallicity were taken from \citet{Bell1981ApJ...248.1031B}, while
the microturbulence  and the abundances of C, N and O are the values found by
\citet{Lambert1982ApJ...254..663L} who have used the model parameters
found by \citet{Bell1981ApJ...248.1031B}. The adopted stellar
parameters for Sirius are thus \teff\ = $10150$\,K, $\log$ g = $4.30$,  \vt\ = $2.0$\,\kms, [Fe/H] = $0.50$,
$\varepsilon$(C) = $7.97$,  $\varepsilon$(N) = $8.15$,
$\varepsilon$(O) = $8.55$, $\pi = 379.21 \pm 1.58$\,mas, resulting in
\ad\ = $6.17 \pm 0.38$\,mas, R = $1.75 \pm 0.11$\,\Rsun, M$_g$ = $2.22
\pm 1.06$\,\Msun\ and L = $29 \pm 6$\,\Lsun. Using these parameters, the
deviation estimating parameters $\beta$ from the Kolmogorov-Smirnov
statistics are $\beta_{\mathrm{1A}}$ = 0.041, $\beta_{\mathrm{1B}}$ = 0.017,
$\beta_{\mathrm{1D}}$ = 0.098 and $\beta_{\mathrm{1E}}$ = 0.027.

Sirius is the star in our sample with the highest gravity. So, it is
not surprising that the synthetic spectrum deviates largely from the observed
spectrum in band 1D, where the Humphreys lines determine the spectral
signature. The pronounced discrepancy seen around 6\,\mic\ is a
consequence of the use of an inaccurate model for the memory-effect
correction in the OLP6.0 calibration of the ISO-SWS data. Consequently the
relative spectral response functions are still not well determined in
band 2.

\subsubsection{Comparison with other published stellar parameters}\label{litersirius}

\begin{table*}[t!]
\caption{\label{litsirius}See caption of Table \ref{litacena}, but now
for Sirius.}
\begin{center}
\setlength{\tabcolsep}{1.2mm}
\tiny
\begin{tabular}{cccccccccccl} \hline
\rule[-3mm]{0mm}{8mm}  \teff & $\log$ g & M & $\xi_t$
&  [Fe/H] & $\varepsilon$(C) & $\varepsilon$(N) & $\varepsilon$(O)
& $\theta_d$ & L & R & Ref.\\ \hline \hline
\rule[-0mm]{0mm}{5mm}
$9697$ & & & & & & & & $6.20$ & & & 1. \\
($9440$) & ($4.33$) & ($2.20$) & & & & & & & ($23.4$) & ($1.68$) & 2. \\
$10150 \pm 400$ & $4.30 \pm 0.20$ & (2.1) & & $0.50 \pm 0.30$ & & & & & &
& 3. \\
($10150$) & ($4.30 \pm 0.20$) & & $2.0 \pm 0.5$ & $0.60 \pm 0.30$
& $7.97 \pm 0.15$ & $8.15 \pm 0.15$ & $8.55 \pm 0.12$ & & & & 4. \\
 & & & & & & & & ${\mathit{5.89}}$ & & $1.675$ & 5. \\
$9900$ & 4.32 & & & & & & & & & & 6. \\
($10000$) & ($4.30$) & & $2.0$ & $0.50$ & & & & & & & 7. \\
10100 & & & & & & & & & ($26.75$) &
$1.69 \pm 0.05$ & 8. \\
$9900$ & $4.30$ & & $2.0$ & & $7.85 \pm 0.06$ & & & & & & 9. \\
$9870$ & $4.32$ & & $1.7 \pm 0.2$ & $0.28$ & $7.82$ & $ > 8.20$ & & & & & 10. \\
$9940 \pm 210$ & ($4.33$) & ($2.20$) & & & & & & & & & 11$^{a}$. \\
$9940 \pm 210$ & $4.20 \pm 0.15$ & ($2.20$) & & & & & & & & & 11$^{b}$. \\
$9940 \pm 210$ & $4.3 \pm 0.5$ & ($2.20$) & & & & & & & & & 11$^{c}$. \\
$9870 \pm 200$ & $4.40 \pm 0.14$ & & $2.0$ & $0.49 \pm 0.29$ & & & &
& & & 12. \\
$9900$ & $4.30$ & & $2.0$ & $0.64$ & $8.03 \pm 0.54$ & & & & & & 13. \\
$9945 \pm 122$ & & & & & & & & ${\mathit{5.92 \pm 0.09}}$ & & & 14$^{a}$.\\
$9943$ & & & & & & & & ${\mathit{5.86}}$ & & &  14$^{b}$.\\
($9880$) & $4.40 \pm 0.20$ & & $1.85 \pm 0.30$ & $0.50$ & $7.64
\pm 0.06$ & $8.04 \pm 0.15$ & $8.63 \pm 0.05$ & & & & 15. \\
\rule[-3mm]{0mm}{3mm}($10150$) & ($4.30$) & $2.22 \pm 1.06$ & ($2.0$)
& ($0.50$) & ($7.97$) & ($8.15$) & ($8.55$) & $6.17 \pm 0.38$ & $29 \pm 6$ &
$1.75 \pm 0.11$ & 16. \\
\hline
\end{tabular}
\end{center}
{\footnotesize{
1.\ \citet{Blackwell1980A&A....82..249B};
2.\ \citet{Popper1980ARA&A..18..115P};
3.\ \citet{Bell1981ApJ...248.1031B};
4.\ \citet{Lambert1982ApJ...254..663L};
5.\ \citet{Moon1985Ap&SS.117..261M};
6.\ \citet{Moon1985MNRAS.217..305M};
7.\ \citet{Sadakane1989PASJ...41..279S};
8.\ \citet{Volk1989AJ.....98.1918V};
9.\ \citet{Lemke1990A&A...240..331L};
10.\ \citet{Hill1993A&A...276..142H};
11.\ \citet{Smalley1995A&A...293..446S};
12.\ \citet{Hui-Bon-Hoa1997A&A...323..901H};
13.\ \citet{Rentzsch-Holm1997A&A...317..178R};
14.\ \citet{DiBenedetto1998A&A...339..858D};
15.\ \citet{Qiu2001Apj...548...953Q};
16.\ present results}}
\end{table*}

\begin{itemize}
\item{\underline{Assumed parameters:}\\
Several authors determined the effective temperature and the surface
gravity for Sirius by using $uvby\beta$ photometry and the grids of
\citet{Moon1985MNRAS.217..305M} \citep[e.g.,
][]{Lemke1990A&A...240..331L, Hill1993A&A...276..142H,
Hui-Bon-Hoa1997A&A...323..901H}. The values of the effective
temperature obtained in that way are in between the values derived
from the IRFM \citep[e.g., ][]{Blackwell1980A&A....82..249B} and the
values derived from comparing observed and
theoretical fluxes \citep[e.g., ][]{Bell1981ApJ...248.1031B}, with the
IRFM-results being the lowest ones. The
derived values for the gravity from the Balmer line profile and from
photometric data are in good agreement, and more or less the same
abundance pattern is derived by different authors. The microturbulence
and C, N and O abundance deduced by
\citet{Lambert1982ApJ...254..663L} were assumed as input parameters,
who have taken \teff, $\log$ g and [Fe/H] from
\citet{Bell1981ApJ...248.1031B} as model parameters.}
\item{\underline{Deduced parameters:}\\
The only indirect
measurement of the angular diameter available is from IRFM. Our
derived angular diameter of \ad$ = 6.17 \pm 0.38$\,mas
corresponds with the value of
\citet{Blackwell1980A&A....82..249B}. Contrary to the target $\alpha$
Car, a good agreement is found for the luminosity and radius values of
Sirius between our deduced values and the ones of
\citet{Volk1989AJ.....98.1918V}. The reason is situated in the
Hipparcos' parallax now being almost the same as the parallax of
\citet{Hoffleit1982} ($\pi = 378$\,mas), used to determine R and L
from \ad. \citet{Moon1985Ap&SS.117..261M} used a quite different
method, based on a new relation between the visual surface brightness
$F_{\nu}$ and the $(b-y)_0$ colour index of $uvby\beta$ photometry
(see the appendix available at
http://www.ster.kuleuven.ac.be/\~\,$\!$leen/artikels/\\ISO3/appendix.ps).
His quoted value lies within the error bars of our deduced
value.

Not only for Sirius, but also for other warm stars in the sample,
we note a large error bar on the derived mass M$_g$, which mainly
depends on the error in the gravity. This demonstrates that other
methods for mass determination (e.g.\ from data of eclipsing and
visual binaries) are far more useful than the
underlying method for the M$_g$ determination (from the gravity and the
radius).
}
\end{itemize}


\subsection{Vega\rm{: AOT01, speed 3, revolution 178}} \label{disvega}

\subsubsection{Some specific calibration details}

This speed-3 observation of Vega in revolution 178 had some problems
with the pointing: dy = $-0.608''$ and dz = $-1.179''$. Switching then
to a larger aperture between band 1B and band 1D results in a
flux-jump, which is clearly visible in the factors used
to shift the different sub-bands. In order to have a smooth spectrum,
we had to multiply the data of bands 1A and 1B by a factor 1.06
\citepalias[see Table 3 in][]{Decin2000b}.

\subsubsection{Comparison between the ISO-SWS spectrum and the synthetic
spectrum (Fig.\ \ref{vega})}

\begin{figure}[h!]
\begin{center}
\resizebox{0.5\textwidth}{!}{\rotatebox{90}{\includegraphics{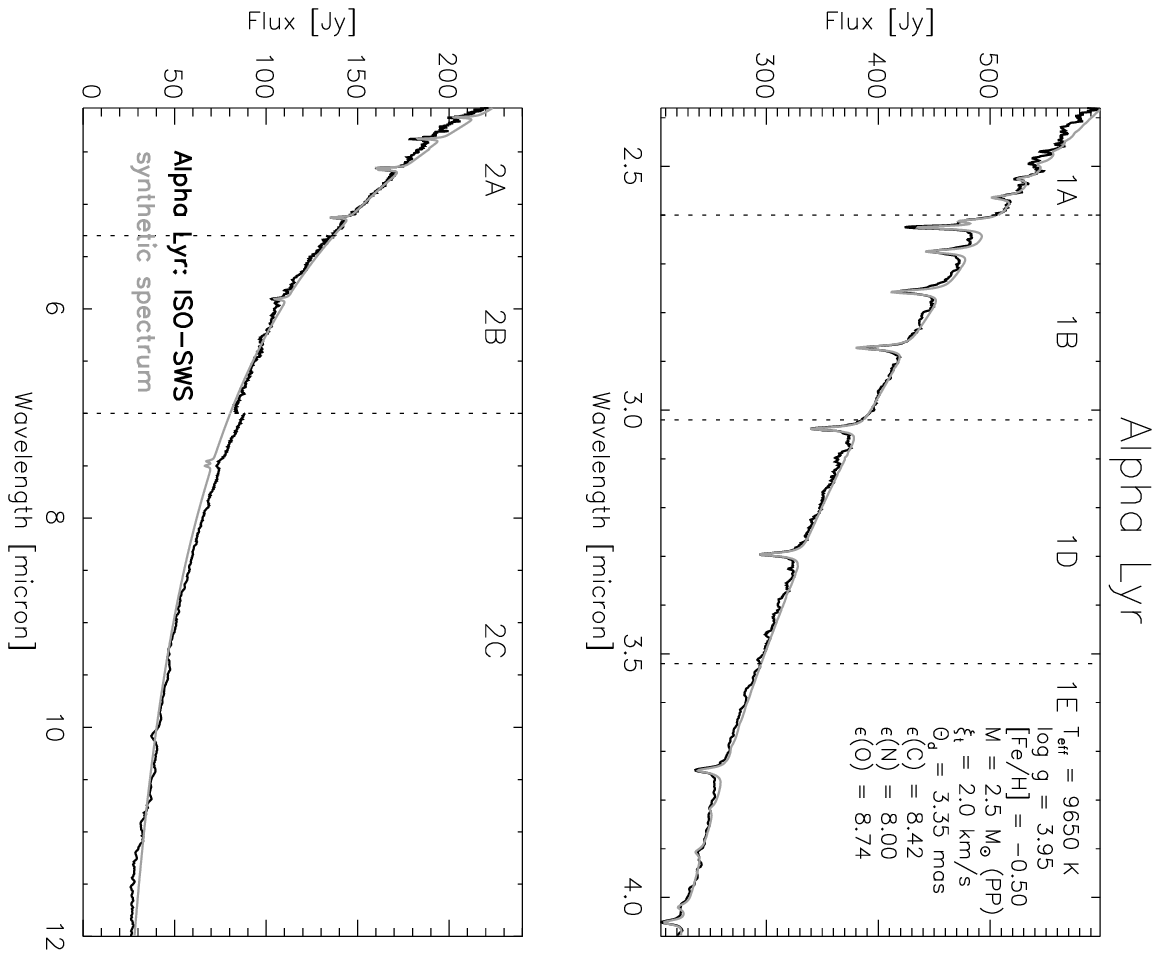}}}
\caption{\label{vega} Comparison between band 1 and band 2 of the
ISO-SWS data of $\alpha$ Lyr (black) and the synthetic spectrum
(grey) with stellar parameters \teff\ = 9650\,K, $\log$ g = 3.95, M
= 2.5\,\Msun, [Fe/H] = $-0.50$, \vt\ = 2.0\,\kms,
$\varepsilon$(C) = 8.42, $\varepsilon$(N) = 8.00, $\varepsilon$(O)
=  8.74 and \ad\ = 3.35\,mas. A coloured version of this plot is
available in the Appendix as Fig.\ \ref{vegacol}.}
\end{center}
\end{figure}

As will be discussed in Sect.\ \ref{litervega}, the following stellar
parameters were adopted for Vega: \teff = 9650\,K, $\log$ g = 3.95,
\vt\ = 2.0\,\kms, [Fe/H] = $-0.50$, $\varepsilon$(C)$ = 8.42$,
$\varepsilon$(N)$ = 8.00$, $\varepsilon$(O)$ = 8.74$.
From the ISO-SWS
spectrum of Vega, an angular diameter of $3.35 \pm 0.20$\,mas was
deduced, with then yields a stellar radius of $2.79 \pm 0.17$\,\Rsun, a
gravity-inferred mass of $2.54 \pm 1.21$\,\Msun\ and a stellar luminosity
of $61 \pm 9$\,\Lsun. Both the synthetic
spectrum based on these parameters and  the ISO-SWS spectrum of Vega
are displayed in Fig.\ \ref{vega}. The corresponding $\beta$-values
are $\beta_{\mathrm{1A}}$ = 0.057, $\beta_{\mathrm{1B}}$ = 0.046,
$\beta_{\mathrm{1D}}$ = 0.010, $\beta_{\mathrm{1E}}$ = 0.041.

In spite of the lower signal-to-noise and lower resolution of the ISO-SWS
observation, good $\beta$-values are obtained. One indeed would not
expect such a low $\beta$-value in band 1D, the wavelength-range in
which the hydrogen Hymphreys lines are absorbing. The reason for this
is twofold: first of all, Vega has the lowest gravity in
our sample of
main-sequence stars, resulting in a smaller pressure-broadening and
thus in smaller hydrogen lines. Consequently, the discrepancy with the
synthetic predictions, which underestimate the strength of the
Humphreys lines, is not as pronounced as for the other
main-sequence stars in our sample. Secondly, Vega has been observed by
using the AOT01 speed-3 option and we have already pointed out the
small --- but visible in the spectrum --- mispointing for this
observation. The larger noise inherent to this observation can
therefore partly camouflage the problem with the theoretical
computation of the hydrogen Humphreys lines. Our statistical test will
not report this problem, since the Kolmogorov-Smirnov test is a {\it
global} goodness-of-fit test and fitting by eye was still necessary
to detect this kind of problems. A new
statistical approach in which a {\it global} and {\it local}
goodness-of-fit test are combined is therefore now under development.

\subsubsection{Comparison with other published stellar parameters}\label{litervega}

\begin{table*}[htb]
\caption{\label{litvega}See caption of Table \ref{litacena}, but now
for Vega.}
\begin{center}
\setlength{\tabcolsep}{1.3mm}
\tiny
\begin{tabular}{cccccccccccl} \hline
\rule[-3mm]{0mm}{8mm}  \teff & $\log$ g & M & $\xi_t$
&  [Fe/H] & $\varepsilon$(C) & $\varepsilon$(N) & $\varepsilon$(O)
& $\theta_d$ & L & R & Ref.\\ \hline \hline
\rule[-0mm]{0mm}{5mm}$9468$ & & & & & & & & 3.35 & & & 1. \\
$9650$ & $3.90 \pm 0.20$ & ($2.0$) & & $-0.41 \pm 0.30$ & & &  &
${\mathit{3.24 \pm 0.07}}$ & & $2.83$ & 2. \\
($9660$) & ($3.94$) & & $2.0$ & $-0.60$ & & & & & & & 3. \\
($9650$) & ($3.90$) & & $2.0 \pm 0.5$ & $-0.40 \pm 0.30$
& $8.57 \pm 0.15$ & $7.93 \pm 0.15$ & $8.82 \pm 0.12$ & & & & 4. \\
 & & & & & & & & & &$2.588$ & 5$^a$. \\
 & & & & & & & & & &$2.234$ & 5$^b$. \\
$9500$ & $3.90$ & & & & & & & & & & 6. \\
($9500$) & ($3.90$) & & 2.0 & $-0.55 \pm 0.10$ & & & & & & & 7. \\
 & & & & & & & & & ($62.66$) & $2.83 \pm
0.13$ & 8. \\
($9400$) & ($3.95$) & & 0.6 & $-0.56 \pm 0.15$ & $8.19$ & & & &
& & 9. \\
$9500$ & $3.90$ & & & & $8.49$ & & & & & & 10. \\
($9650$) & ($3.95$) & & ($2.0$) & $-0.53 \pm 0.15$ &
$8.42 \pm 0.15$ & $8.00 \pm 0.15$ & $8.74 \pm 0.15$ & & & & 11. \\
$9560$ & $4.05$ & & $2.0 \pm 0.2$ & $-0.54$ & $8.47$ & $>8.40$ & & & & & 12. \\
9600 & & & & & & & & & & & 13. \\
$9450$ & $4.00$ & & ($2.0$) & $-0.56 \pm 0.05$ & & & & & & & 14. \\
$9550 \pm 50$ & $3.95 \pm 0.05$ & & ($2.0$) & $-0.50$ & & & & &
& & 15. \\
$9600 \pm 180$ & $4.00 \pm 0.10$ & & & & & & & & & & 16$^a$. \\
$9600 \pm 180$ & $3.80 \pm 0.30$ & & & & & & & & & & 16$^b$. \\
$9830 \pm 320$ & & & & & & & & & & & 17$^a$.\\
$9660$ & & & & & & & & & & & 17$^b$. \\
$9660 \pm 140$ & & & & & & & & ${\mathit{3.24 \pm 0.07}}$ & & & 18$^a$. \\
$9469$ & & & & & & & & ${\mathit{3.20}}$ & & & 18$^b$. \\
$9553 \pm 111$ &  & & & & & & & ${\mathit{3.28 \pm 0.01}}$ & & & 19. \\
($9430$) & $3.95 \pm 0.20$ & & $1.50 \pm 0.30$ & $-0.57$ & $8.46
\pm 0.13$ & $8.00 \pm 0.02$ & $9.01 \pm 0.14$ & & & & 20. \\
\rule[-3mm]{0mm}{3mm} ($9650$) & ($3.95$) & $2.54 \pm 1.21$ & ($2.0$)
& ($-0.50$) &
($8.42$) & ($8.00$) & ($8.74$) & $3.35 \pm 0.20$ & $61 \pm 9$ & $2.79
\pm 0.17$ & 21. \\
\hline
\end{tabular}
\end{center}
{\footnotesize{
1.\ \citet{Blackwell1980A&A....82..249B};
2.\ \citet{Dreiling1980ApJ...241..736D};
3.\ \citet{Sadakane1981PASJ...33..189S};
4.\ \citet{Lambert1982ApJ...254..663L};
5.\ \citet{Moon1985Ap&SS.117..261M};
6.\ \citet{Moon1985MNRAS.217..305M};
7.\ \citet{Gigas1986A&A...165..170G};
8.\ \citet{Volk1989AJ.....98.1918V};
9.\ \citet{Adelman1990ApJ...348..712A};
10.\ \citet{Lemke1990A&A...240..331L};
11.\ \citet{Venn1990ApJ...363..234V};
12.\ \citet{Hill1993A&A...276..142H};
13.\ \citet{Napiwotzki1993A&A...268..653N};
14.\ \citet{Smith1993A&A...274..335S};
15.\ \citet{Castelli1994A&A...281..817C};
16.\ \citet{Smalley1995A&A...293..446S};
17.\ \citet{Sokolov1995A&AS..110..553S};
18.\ \citet{DiBenedetto1998A&A...339..858D};
19.\ \citet{Ciardi2001ApJ...559.1147C};
20.\ \citet{Qiu2001Apj...548...953Q};
21.\ present results}}
\end{table*}

The bright star Vega has been studied extensively in recent years
because it serves as the primary standard star for photoelectric
spectrophotometry. Since so many publications are available for this
star --- as well as for Sirius --- we only have quoted the main
publications in the last two decades.
\begin{itemize}
\item{\underline{Assumed parameters:}\\
Inspecting Table \ref{litvega},
we can see that the published values for the different parameters all
do agree well. As for Sirius, the temperature values derived from IRFM
are somewhat lower than the other published values (see also the
remark made by \citet{Napiwotzki1993A&A...268..653N}, who quoted that
the IRFM temperatures are too low by 1.6 -- 2.8\,\% for main-sequence
stars). A detailed study of Vega was made by
\citet{Dreiling1980ApJ...241..736D}. From this study
\citet{Venn1990ApJ...363..234V} have adopted the effective temperature
(\teff = $9650$\,K) and the gravity ($\log$ g = 3.95). Using the
microturbulent velocity found by
\citet{Lambert1982ApJ...254..663L} (\vt\ = $2.0$\,\kms),
\citet{Venn1990ApJ...363..234V} have determined the chemical
composition for Vega.  Using the solar metallicity obtained from
meteoritic data, $\varepsilon$(Fe)$ = 7.51$, as a reference value, their abundances for carbon,
nitrogen, oxygen and iron were respectively: $\varepsilon$(C)$ = 8.42
\pm 0.15$, $\varepsilon$(N)$ = 8.00 \pm 0.15$, $\varepsilon$(O)$ =
8.74 \pm 0.15$ and [Fe/H]$ = -0.53 \pm 0.15$. }
\item{\underline{Deduced parameters:}\\
Our angular diameter derived from the ISO-SWS spectrum
corresponds to the IRFM value from
\citet{Blackwell1980A&A....82..249B}. The same note as made in Sect.\
\ref{litersirius} concerning the luminosity and radius values
mentioned by \citet{Volk1989AJ.....98.1918V} can be made: since the
parallax value mentioned by \citet{Hoffleit1982} ($\pi = 133$\,mas)
only differs with the Hipparcos' value by a factor 1.03, the values for
R and L are in close agreement. This can also be said for the
radius value given by \citet{Dreiling1980ApJ...241..736D}, who too
have used the parallax of \citet{Hoffleit1982}. The new relation
established by \citet{Moon1985Ap&SS.117..261M} results in stellar
radius values which are somewhat lower than our deduced value and the
values mentioned by \citet{Moon1985Ap&SS.117..261M} and
\citet{Volk1989AJ.....98.1918V}. }
\end{itemize}


\section{Conclusion}

The five warmest stars in a sample of 16 stars --- used for the
calibration of the detectors of ISO-SWS --- have been discussed
spectroscopically. The absence of molecular features and the presence
of atomic features whose oscillator strengths are not well-known
rendered the determination of the effective temperature, gravity,
microturbulent velocity, metallicity and the abundance of C, N, and O from the
ISO-SWS data unfeasible. Good-quality published values were then used
for the computation of the synthetic spectra. In general, no more
discrepancies than the ones reported in \citet{Decin2000b} have been
detected. A comparison with other --- lower resolution ---
ISO-SWS data revealed a rather good relative agreement ($\sim
2$\,\%), but the absolute flux-level and so the deduced angular
diameter could differ by up to 16\,\%. Nevertheless, the angular
diameter, luminosity and stellar radius deduced from the ISO-SWS data
are in good agreement with other published values deduced from other
data and/or methods.

Since this research has shown clearly that the available oscillator strengths of
atomic transitions in the infrared are at the moment still very
inaccurate, one of us (J.\ S.) has worked on a new atomic linelist by deducing
new oscillator strengths from the high-resolution ATMOS spectrum of
the Sun (625 -- 4800 cm$^{-1}$) \citep{Sauval2000}. This new atomic
linelist will be presented in Paper~V of this series.


\begin{acknowledgements}LD acknowledges support from the Science Foundation of
Flanders. This research has made use of the SIMBAD database, operated
at CDS, Strasbourg, France and of the VALD database, operated at Vienna,
Austria. It is a pleasure to thank the referees, J.\ Hron and F.\
Kupka, for their careful reading of the manuscript and for their
valuable suggestions.
\end{acknowledgements}


\aareferences
